\documentclass[twocolumn]{aastex63}

\usepackage{amsmath}
\usepackage{amssymb}
\usepackage{amsthm}
\usepackage{natbib,aasdefs,url,bm}
\usepackage{graphicx}
\usepackage{color}
\usepackage{mathtools,  epsfig, lipsum}
\usepackage{subfigure}
\usepackage{textcomp}
\usepackage{multirow,booktabs,colortbl,tabularx}
\usepackage{enumitem}
\usepackage{booktabs}
\usepackage{hyperref}

\shorttitle{Neutrinos and Gamma-rays from Shock-Powered Transients}
\shortauthors{Fang, Metzger, Vurm, Aydi, Chomiuk }

\begin{document}

\title{High-Energy Neutrinos and Gamma-Rays from Non-Relativistic Shock-Powered Transients}

\author{Ke Fang} 
\affil{Kavli Institute for Particle Astrophysics and Cosmology (KIPAC), Stanford University, Stanford, CA 94305, USA}
\affil{NHFP Einstein Fellow}

\author{Brian D. Metzger}
\affil{Department of Physics and Columbia Astrophysics Laboratory, Columbia University,\\ Pupin Hall, New York, NY 10027, USA}
\affil{Center for Computational Astrophysics, Flatiron Institute, 162 5th Ave, New York, NY 10010, USA} 

\author{Indrek Vurm}
\affil{Tartu Observatory, Tartu University, 61602 T{$\bar o$}ravere, Tartumaa, Estonia}

\author{Elias Aydi}
\affil{Center for Data Intensive and Time Domain Astronomy, Department of Physics and Astronomy, Michigan State University, East Lansing, MI, USA}

\author{Laura Chomiuk}
\affil{Center for Data Intensive and Time Domain Astronomy, Department of Physics and Astronomy, Michigan State University, East Lansing, MI, USA}

\begin{abstract}
Shock interaction has been argued to play a role in powering a range of optical transients, including supernovae, classical novae, stellar mergers, tidal disruption events, and fast blue optical transients.  These same shocks can accelerate relativistic ions, generating high-energy neutrino and gamma-ray emission via hadronic pion production.  The recent discovery of time-correlated optical and gamma-ray emission in classical novae has revealed the important role of radiative shocks in powering these events, enabling an unprecedented view of the properties of ion acceleration, including its efficiency and energy spectrum, under similar physical conditions to shocks in extragalactic transients.  

Here we introduce a model for connecting the radiated optical fluence of non-relativistic transients to their maximal neutrino and gamma-ray fluence.  We apply this technique to a wide range of extragalactic transient classes in order to place limits on their contributions to the cosmological high-energy gamma-ray and neutrino backgrounds.  Based on a simple model for diffusive shock acceleration at radiative shocks, calibrated to novae, we demonstrate that several of the most luminous transients can accelerate protons up to $10^{16}$ eV, sufficient to contribute to the IceCube astrophysical background.  Furthermore, several of the considered sources$-$particularly hydrogen-poor supernovae$-$may serve as ``gamma-ray- hidden" neutrino sources due to the high gamma-ray opacity of their ejecta, evading constraints imposed by the non-blazar {\it Fermi}-LAT background.  However, adopting an ion acceleration efficiency $\sim 0.3-1\%$ motivated by nova observations, we find that currently known classes of non-relativistic, potentially shock-powered transients contribute at most a few percent of the total IceCube background.  
\end{abstract}


\section{Introduction}

Optical time-domain surveys have in recent years discovered new classes of explosive transients characterized by a wide diversity of properties (e.g.~\citealt{Villar+17}).  These include exotic channels of massive star death, such as ``superluminous supernovae" (SLSNe; \citealt{GalYam19,Inserra19}) of both hydrogen-rich \citep{Smith+07} and hydrogen-poor \citep{Quimby+11} varieties; tidal disruption events of stars by massive black holes (TDEs; \citealt{Gezari+12,Stone+19}); ``luminous red novae" (LRNe; e.g.~\citealt{Tylenda+11}) and dusty infrared-bright transients \citep{Kasliwal+17} from merging binary stars; and ``fast blue optical transients" (FBOTs; e.g.~\citealt{Drout+14}) of an uncertain origin likely related to massive star death.  

Many of these events reach peak luminosities which are greater than can be understood by the traditional energy sources available to supernovae, such as radioactive decay or the initial heat generated during the dynamical explosion, merger, or disruption.  An additional, internal power source is clearly at play.  One of the most promising ways of enhancing the optical output from a transient are via shocks, generated as the explosion ejecta (or streams of stellar debris in the case of TDEs) collides with themselves or an external medium.  For a wide large range of conditions these shocks are {\it radiative}, meaning that due to the high gas densities the thermal cooling time behind the shock is short compared to the expansion time.  Under these conditions the shocked gas emit copious UV/X-ray emission which is absorbed with high efficiency by surrounding gas and ``reprocessed" downwards into the visual waveband, enhancing or even dominating the transient light (e.g.~\citealt{Chevalier&Fransson94}).

Shock interaction is commonly invoked to power the light curves of SLSNe (e.g.~\citealt{Smith&McCray07,Chevalier&Irwin11,Moriya+14,Sorokina+16}), particularly the hydrogen-rich variety (SLSNe-II) in which narrow emission lines directly reveal the presence of dense slow gas ahead of the ejecta (dubbed ``Type IIn" when the hydrogen lines are narrow; \citealt{Schlegel90}).  However, embedded shock interaction could also power SN light curves even in cases where emission features or other shock signatures are not visible, for example when an compact circumstellar disk is overtaken by faster opaque ejecta (e.g.~\citealt{Andrews&Smith18}).  Shells or outflows of dense external gas surrounding supernovae can be the result of intense mass-loss from the star in the years and decades prior to its explosion \citep{Smith14}.  In the case of extremely massive, metal-poor stars, this can include impulsive mass ejection as a result of the pulsational pair instability \citep{Woosley+07,Tolstov+16}.  

Similarly in binary star mergers, shock interaction can take place between fast matter ejected during the dynamical ``plunge" phase at the end of the merger process and slower outflows from the earlier gradual inspiral \citep{Pejcha+17,MacLeod+18}; these embedded shocks may be responsible for powering the plateau or secondary maxima observed in the light curves of LRN \citep{MetzgerPejcha17}.  Shock-mediated collisions between the bound streams of the disrupted star in TDEs may power at least part of the optical emission in these events \citep{Piran+15,Jiang+16}.  The optical emission from FBOTs, such as the nearby and well-studied AT2018cow \citep{Prentice+18,Perley+19}, could also be powered by internal shock interaction in explosions with a low ejecta mass \citep{Margutti+19,Tolstov+19,Piro&Lu20}.\footnote{However, note that an energetic compact object$-$a newly-born magnetar or accreting black hole$-$provides an alternative energy source in FBOTs and SLSNe \citep{Kasen&Bildsten10,Woosley10}, which could also be a source of neutrinos \citep{Fang+19}.} 

In each of the extragalactic transients cited above, the inference of shock interaction is at best indirect.  However, a direct confirmation of embedded shock-powered emission has become possible recently from a less energetic (but comparatively nearby) class of Galactic transients: the classical novae.  Over the past decade, the {\it Fermi} Large Area Telescope (LAT) has detected $\sim 0.1-10$ GeV gamma-ray emission coincident with the optical emission from over 10 classical novae \citep{Ackermann+14,Cheung+16,Franckowiak+18}.  The non-thermal gamma-rays are generated by relativistic particles accelerated at shocks (via the diffusive acceleration process; \citealt{Blandford&Ostriker78,Eichler79,Bell04}), which arise due to collisions internal to the nova ejecta \citep{Chomiuk+14,Metzger+14}.  

Non-thermal gamma-ray emission in novae could in principle be generated either by relativistic electrons (which Compton up-scatter the nova optical light or emit bremsstrahlung radiation in the GeV band$-$the ``leptonic" mechanism) or via relativistic ions colliding with ambient gas (generating pions which decay into gamma-rays$-$the ``hadronic" mechanism).  However, several arguments favor the hadronic mechanism and hence the presence of ion acceleration at nova shocks.  For example, strong magnetic fields are required near the shocks to confine and accelerate particles up to sufficiently high energies $\gtrsim 10-100$ GeV to generate the observed gamma-ray emission; embedded in the same magnetic field, however, relativistic electrons lose energy to lower-frequency synchrotron radiation faster than it can be emitted as gamma-rays, disfavoring the leptonic models \citep{Li+17,Vurm&Metzger18}.

The ejecta surrounding the shocks in novae are sufficiently dense to act as a ``calorimeter" for converting non-thermal particle energy into gamma-rays \citep{Metzger+15}.  For similar reasons of high densities, the shocks are radiative and their power is reprocessed into optical radiation with near-unity efficiency \citep{Metzger+14}.  Stated another way, both the thermal and non-thermal particles energized at the shocks find themselves in a fast-cooling regime.  As a result, the gamma-ray and shock-powered optical emission should trace one another and the ratio of their luminosities can be used to directly probe the particle acceleration efficiency \citep{Metzger+15}.  In two novae with high-quality gamma-ray light curves, ASASSN16ma \citep{Li+17} and V906 Car \citep{Aydi+20}, the time-variable optical and gamma-ray light curves are observed to track each other, confirming predictions that radiative shocks can power the optical emission in novae \citep{Metzger+14}.  

Applying the above technique, one infers an efficiency of non-thermal particle acceleration in novae of $\epsilon_{\rm rel} \sim 0.3-1\%$ \citep{Li+17,Aydi+20}.  This is low compared to the $\epsilon_{\rm rel} \sim 10\%$ efficiency one finds for the adiabatic shocks in supernova remnants (e.g.~\citealt{Morlino&Caprioli12}) or the maximal value $\epsilon_{\rm rel} \sim 20\%$ found from particle-in-cell simulations of diffusive shock acceleration for the optimal case in which the upstream magnetic field is quasi-parallel to the shock normal \citep{Caprioli&Spitkovsky14}.  In novae$-$as in other shock-powered transients$-$the magnetic field of the upstream medium is generically expected to be wrapped in the toroidal direction around the rotation axis of the outflow (``Parker spiral"; \citealt{Parker58}), perpendicular to the radial shock direction and hence in the quasi-perpendicular regime for which little or no particle acceleration is theoretically predicted \citep{Caprioli&Spitkovsky14}.  The small efficiency $\sim 0.3-1\%$ that nevertheless is obtained may arise due to the irregular, corrugated shape of the radiative-shock front, which allows local patches of the shock to possess a quasi-parallel shock orientation and hence to efficiently accelerate particles \citep{Steinberg&Metzger18}.

Gamma-rays generated from the decay of $\pi^{0}$ in hadronic accelerators are accompanied by a similar flux of neutrinos from $\pi^{\pm}$ decay.  A future detection of $\sim$ GeV-TeV neutrino emission, likely from a particularly nearby nova, would thus serve as a final confirmation of the hadronic scenario \citep{Razzaque+10,Metzger+16}.  However, compared to supernovae, the relatively low kinetic energies of classical novae make them sub-dominant contributors to the cosmic-ray or neutrino energy budget in the Milky Way or other galaxies.  On the other hand, with the exception of their luminosities, many of the physical conditions which characterize nova shocks (gas density, evolution timescale) are broadly similar to those of more energetic extragalactic transients.  The advantage of novae$-$being among the brightest transients in the night sky$-$is their relative proximity, which enables a detailed view of their gamma-ray emission and hence particle acceleration properties.  

For comparison, non-thermal gamma-rays have not yet been detected from extragalactic supernovae in either individual or stacked analysis (\citealt{2015ApJ...807..169A, RenaultTinacci+18,Murase+19}, with a few possible exceptions; \citealt{Yuan+18,Xi+20}).  This is despite the potential for shock interaction within these sources$-$if prevalent$-$to be major contributors of high-energy cosmic rays, gamma-rays, and neutrinos (e.g.~\citealt{Murase+11,Katz+11,Chakraborti+11,2013ApJ...769L...6K, 2014MNRAS.440.2528M, 2016APh....78...28Z, Marcowith+18,2018PhRvD..97h1301M, Zhang&Murase19,Cristofari+20}).

In this paper we apply the knowledge of particle acceleration at radiative shocks, as gleaned from recent studies of classical novae \citep{Li+17,Aydi+20}, to assess the prospects of interacting supernovae and other non-relativistic, shock-powered extragalactic transients as sources of high-energy gamma-ray emission and neutrinos.  An astrophysical neutrino population above $\sim 10$~TeV has been measured by the IceCube Observatory \citep{IceCube+13, Schneider19,Stettner19}. 
The sources that contribute to the bulk of high-energy neutrinos remain unknown  \citep{IceCube+19,IceCube+20}, though hints of sources have been suggested \citep{IceCube+18b, IceCube+18, IceCube+20}.  We are thus motivated to consider to what extent shock-powered transients, under optimistic but realistic (i.e.~observationally-calibrated) assumptions, are capable of contributing to the neutrino background.  

Intriguingly, the magnitude of IceCube's diffuse neutrino flux is comparable to that of the {\it Fermi}-LAT isotropic $\gamma$-ray background (IGRB) around $\sim 100$~GeV \citep{2015ApJ...799...86A, 2015PhRvD..91l3001D}, and to avoid over-producing the IGRB the neutrino sources were suggested to be ``hidden", i.e.~locally opaque to  1-100~GeV $\gamma$-rays (e.g.  \citealt{2001APh....15...87B, 2016PhRvL.116g1101M, 2020PhRvD.101j3012C, 2020arXiv200707911C}).  Given the high column densities of shock-powered transients, they offer one of only a handful of potentially gamma-ray-hidden neutrino sources, further motivating our study.

This paper is organized as follows.  In $\S\ref{sec:overview}$ we introduce a simple model for non-relativistic shock-powered transients and describe the connection between their high energy gamma-ray/neutrino and optical emissions, as probed via the calorimetric technique.  In $\S\ref{sec:novae}$ we apply the methodology to classical novae and show how observations (particularly modeling of their gamma-ray spectra) can be used to calibrate uncertain aspects of the acceleration process in radiative shocks.   In $\S\ref{sec:zoo}$ we apply the calorimetric technique to place upper limits on the high-energy neutrino and gamma-ray background from the ``zoo" of (potentially) shock-powered transients across cosmic time and compare them to constraints from IceCube and {\it Fermi}.  In $\S\ref{sec:conclusions}$ we summarize our conclusions.

\section{Shock-Powered Supernovae as Cosmic Ray Calorimeters}
\label{sec:overview}

This section introduces a simplified, but also fairly generic, model of shock-powered transients and the general methodology for using their optical light curves to constrain their high-energy gamma-ray and neutrino emission (see Fig.~\ref{fig:cartoon} for a schematic illustration).  In places where specificity is necessary, we focus on the particular case of interaction-powered SNe.  However, most of the conditions derived are broadly applicable to any transient (e.g., novae, TDEs, stellar mergers) in which a non-relativistic shock is emerging from high to low optical depths.  Insofar as possible, we express our results exclusively in terms of observable quantities such as the optical rise time, peak luminosity, or characteristic expansion velocity (measurable, e.g., from optical spectroscopy).  

\subsection{Shock Dynamics and Thermal Emission}

\begin{figure*}
\includegraphics[width= \linewidth] {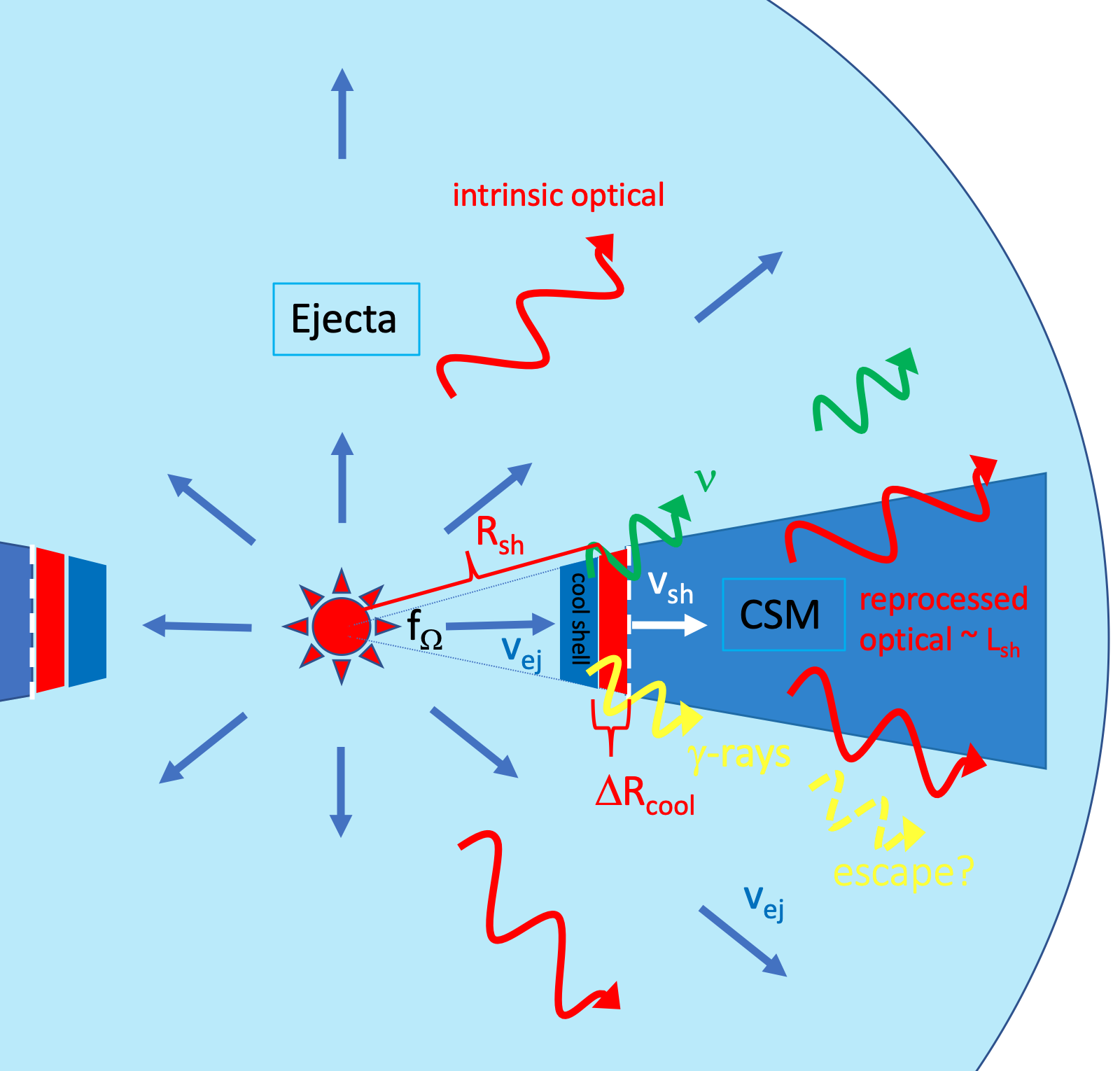} 
\caption{\label{fig:cartoon}
Schematic diagram illustrating the generic scenario for shock-powered emission from explosive non-relativistic transients.  The explosion ejecta collides with a dense external medium (e.g.~circumstellar medium; CSM) of radial density profile $n(r)$ and effective wind-mass loss rate parameter $A \equiv \dot{M}/(4\pi v_w)$ which covers a fractional solid angle $f_{\Omega} < 1$.  The ejecta of mean velocity $\bar{v}_{\rm ej}$ collides with the CSM, driving a shock into the latter with a velocity $v_{\rm sh}$ and kinetic luminosity $L_{\rm sh}$.  UV/X-ray emission from the thin cooling layer behind the shocks is absorbed and reprocessed by the surrounding gas into optical radiation of luminosity $L_{\rm opt} \approx L_{\rm sh}$.  The shock also accelerates relativistic ions which collide with background ions, generating $\pi^{0}(\pi^{\pm})$ which decay into gamma-rays and neutrinos, respectively.  The optical light curve peaks, and the bulk of particle acceleration occurs, when the optical depth surrounding the shock first obeys the condition $\tau_{\rm opt} \lesssim c/v_{\rm sh}$, similar to that required for the formation of a collisionless shock capable of particle acceleration.  At this epoch of peak emission, both thermal particles (which emit via free-free emission) and non-thermal particles (undergoing p-p interactions) are radiative, such that the emitted non-thermal gamma-ray/neutrino emission is proportional to the shock-powered optically-radiated energy.  The thickness of the post-shock region as set by thermal cooling, $\Delta R_{\rm cool}$, is much smaller than the shock radius $R_{\rm sh}$, limiting the maximum particle energy achievable via diffusive shock acceleration (eq.~\ref{eq:Emax2}).    }
\end{figure*}

We consider the collision of spherically expanding homologous ejecta of average velocity $\bar{v}_{\rm ej}$ generated during a dynamical explosion with an effectively stationary external medium (the treatment can easily be generalized to a moving upstream or aspherical ejecta, but for non-relativistic expansion this generally introduces only order-unity changes).  The external medium is assumed to possess a nucleon number density $n \equiv \rho/m_p$ (where $\rho$ is the mass density) with a radial profile $n \propto r^{-k}$, where $k \ge 2$ is a power-law index and to be concentrated into a fractional solid angle $f_{\Omega} \le 1$ (e.g., $f_{\Omega} \sim h/r$ if the external medium is concentrated in a thin equatorial disk of vertical scale-height $h$ and aspect ratio $h/r$).  

One convenient parameterization of the density profile is that of a steady wind of mass-loss rate $\dot{M}$ and velocity $v_{\rm w}$ then $n \simeq \dot{M}/(4\pi f_{\Omega} r^{2}v_w m_p) = A/(m_p r^{2})$, where $A \equiv \dot{M}/(4\pi f_{\Omega}v_w)$.  For example, values of $\dot{M} \sim 10^{-4}-1M_{\odot}$ yr$^{-1}$ and $v_{\rm w} \sim 100-1000$ km s$^{-1}$ are typically inferred by modeling interacting supernovae (e.g.~\citealt{Smith14}), corresponding to $A \simeq 1-10^{5}A_{\star}$ for $f_{\Omega} \sim 1$, where $A_{\star} \equiv 5\times 10^{11}$ g cm$^{-2}$ is a fiducial value for $\dot{M} = 10^{-5}M_{\odot}$ yr$^{-1}$ and $v_{\rm w} = 1000$ km s$^{-1}$ \citep{Chevalier&Li00}.  In general, we expect $k \gtrsim 2$, if the value of $\dot{M}$ is increasing approaching the explosion or dynamical event, as may characterize wave-driven mass-loss from massive stars before they explode as supernovae (e.g.~\citealt{Quataert&Shiode12}) or binary star mergers in which the merger is instigated by unstable mass-transfer and mass-loss which rises rapidly approaching the dynamical coalescence phase (e.g.~\citealt{Pejcha+17}).  In such cases where $k > 2$ the effective value of $A(r)$ is a (decreasing) function of radius, though this detail is not important as we are primarily interested in its value near the optical peak, as discussed further below.  

The collision drives a forward shock into the external medium and a reverse shock back into the ejecta.  When the shocks are radiative (the conditions for which will be verified below) the gas behind both shocks rapidly cools and accumulates into a thin central shell, which propagates outwards into the external medium at a velocity $v_{\rm sh}$ equal to that of the forward shock.  
The shocks reach a radius $R_{\rm sh} \approx v_{\rm sh} t$ by a time $t$ after the explosion.  Given the homologous velocity profile of the ejecta (inner layers slower than outer layers; $v_{\rm ej} \propto r$) in many cases of interest the shell is accelerated to a velocity matching that of the ejecta at a similar radius (e.g.~\citealt{MetzgerPejcha17}), reducing the power of the reverse shock relative to the forward shock by the times of interest near the light curve peak.  Although the discussion to follow focuses on the forward shock-dominated case for concreteness, qualitatively similar results apply to the reverse shock-dominated case.   

The kinetic power of the forward shock is given by
\begin{eqnarray}
    L_{\rm sh} = \frac{9\,\pi}{8}f_{\Omega}\, m_p\,n_{\rm sh}\,v_{\rm sh}^3\,R_{\rm sh}^2  
    = \frac{9}{32}\,\dot{M}\frac{v_{\rm sh}^3}{v_{\rm w}} = \frac{9\pi}{8}A f_{\Omega} v_{\rm sh}^{3}, \nonumber \\
\label{eq:Lsh}
\end{eqnarray}
where $n_{\rm sh} \equiv n(R_{\rm sh})$ is the characteristic upstream density ahead of the shock and $f_{\Omega} \le 1$ is again the fractional solid angle subtended by the shocks interaction (Fig.~\ref{fig:cartoon}).  Gas immediately behind the shock is heated to a temperature 
\begin{equation}
    kT_{\rm sh}\simeq\frac{3}{16}\,\mu m_p\,\,v_{\rm sh}^2 \approx 11 v_{8.5}^{2}\,{\rm keV},
\label{eq:Tsh}
\end{equation}
where $v_{8.5} \equiv v_{\rm sh}/(3000$ km s$^{-1})$ and in the second line we have taken $\mu = 0.62$ for the mean molecular weight of fully-ionized gas of solar composition (we would instead have $\mu \simeq 2$ if the upstream medium is composed of hydrogen-poor gas).  The bulk of the shocks' power $\sim L_{\rm sh}$ is emitted at temperatures $\sim k T_{\rm sh}$ (in the X-ray range for typical shock velocities $v_{\rm sh} \gtrsim 10^{3}$ km s$^{-1}$).  However, due to the large photoelectric opacity of the external medium (at the times during peak light when the bulk of the particle acceleration occurs; see below), most of $L_{\rm sh}$ is absorbed and reprocessed via continuum and line emission into optical wavelengths (consistent, e.g., with the non-detection of luminous X-rays from SLSNe near optical peak; \citealt{Levan+13,Ross&Dwarkadas17,Margutti+18}). 

The shock luminosity $L_{\rm sh}$ is only available to contribute to the supernova light curve after a certain time.  To escape to an external observer, reprocessed emission from the vicinity of the forward shock must propagate through the column of the external medium, $\Sigma = \int_{R_{\rm sh}}^{\infty} ndr \sim n_{\rm sh}R_{\rm sh}$.  The reprocessed optical light will emerge without experiencing adiabatic losses provided that the optical photon diffusion timescale $t_{\rm diff} \approx \tau_{\rm opt}(R_{\rm sh}/c)$, where $\tau_{\rm opt} \equiv \Sigma \sigma_{\rm opt}$ and $\sigma_{\rm opt}$ the effective cross section at visual wavelengths, be shorter than the expansion timescale of the shocked gas, $t_{\rm dyn} \sim R_{\rm sh}/v_{\rm sh}$ over which adiabatic losses occur, i.e.
\begin{equation}
    \tau_{\rm opt} \lesssim c/v_{\rm sh},
    \label{eq:diffuse}
\end{equation}
as is satisfied at times
\begin{equation}
t \gtrsim t_{\rm pk} \approx \frac{c }{v_{\rm sh}^{2}n_{\rm sh}\sigma_{\rm opt}} = \frac{\dot{M}\kappa_{\rm opt}}{4\pi f_{\Omega} c v_{\rm w}} = \frac{A \kappa_{\rm opt}}{ c} ,
\label{eq:tpk}
\end{equation}
where $\kappa_{\rm opt} \equiv \sigma_{\rm opt}/m_p$ is the optical opacity.  We label this critical time $t_{\rm pk}$ since it defines the rise time, and often the peak timescale, of the light curve. 

Equation (\ref{eq:tpk}) neglects corrections to $t_{\rm diff}$ due to non-spherical geometry and assumes that the diffusion of reprocessed optical photons outwards through the shocked gas is the rate-limiting step to their escape, as opposed to additional diffusion through the surrounding ejecta.  Although this assumption is justified in many cases, it is clearly violated in certain cases (e.g., highly aspherical ejecta, $f_{\Omega} \ll 1$; very low CSM mass relative to ejecta mass).  Nevertheless, our cavalier approach is justified since the main goal of our analysis is to provide order of magnitude estimates of the shock properties near optical maximum.

For a wide range of shock-dominated transients, $t_{\rm pk}$ sets the rise time of the light curve to its peak luminosity $L_{\rm pk} \approx L_{\rm sh} = (9\pi/8)A f_{\Omega} v_{\rm sh}^{3}$ (eq.~\ref{eq:Lsh}), with $L_{\rm opt} \ll L_{\rm sh}$ at times $t \ll t_{\rm pk}$ and $L_{\rm opt} \approx L_{\rm sh}$ at $t \gtrsim t_{\rm pk}$.  In general $L_{\rm sh}$ (and hence $L_{\rm opt}$) will decrease after $t_{\rm pk}$ because $A(r)$ is decreasing with radius or because $v_{\rm sh}$ is decreasing as the shock sweeps up mass. 

Combining equations (\ref{eq:Lsh}) and (\ref{eq:tpk}) we can express the shock velocity
\begin{equation}
v_{\rm sh} = \left(\frac{8}{9\pi }\frac{L_{\rm pk}\kappa_{\rm opt}}{ ct_{\rm pk}f_{\Omega}}\right)^{1/3},
\label{eq:vsh}
\end{equation}
in terms of the two other ``observables", $L_{\rm pk}$ and $t_{\rm pk}$.  Here we have assumed that 100\% of the transient's optical light is shock powered, $L_{\rm pk} \approx L_{\rm sh}(t_{\rm pk})$, i.e.~neglecting additional contributions to $L_{\rm pk}$ from e.g., radioactivity, initial thermal energy, or a central engine (though the latter can be a source of energizing the ejecta and driving shocks; e.g.~\citealt{Metzger+14b,Kasen+16, 2017ApJ...849..153F, DeCoene+20}).

\subsection{The Calorimetric Technique}

 Remarkably, the conditions (\ref{eq:diffuse}), (\ref{eq:tpk}) on the optical depth to the shock are very similar to that required for the shock discontinuity to be mediated by collisionless plasma processes instead of by radiation (e.g.~\citealt{Colgate74,Klein&Chevalier78,Katz+11}).  Before this point when the optical depth is higher, relativistic particle acceleration is not possible because trapped radiation thickens the shock transition to a macroscopic scale, precluding the particle injection process \citep{Zeldovich&Raizer67,Weaver76,Riffert88,Lyubarskii&Syunyaev82,Katz+11,Waxman&Katz17}.  

This has two implications: (1) efficient relativistic particle acceleration is unlikely to occur in interacting supernovae and other shock-powered transients well prior to the optical peak; (2) if a fixed fraction $\epsilon_{\rm rel}$ of the shock power $L_{\rm sh}$ is placed into relativistic particles (once eq.~$\ref{eq:diffuse}$ is satisfied), the total energy placed into relativistic particles ($E_{\rm rel} \approx \int_{\rm t_{\rm pk}}^{\infty} \epsilon_{\rm rel}L_{\rm sh}dt$) is proportional to the fraction, $f_{\rm sh}$, of the radiated optical fluence of the supernova ($E_{\rm opt} \approx f_{\rm sh}^{-1}\int_{\rm t_{\rm pk}}^{\infty} L_{\rm sh} dt$) which is powered by shocks.  In other words,
\begin{equation}
    E_{\rm rel} \approx f_{\rm sh}\epsilon_{\rm rel} E_{\rm opt}.
\label{eq:Erel}
\end{equation}
As a corollary, since $f_{\rm sh}< 1$ this implies that $\epsilon_{\rm rel}E_{\rm opt}$ is an upper limit on the energy of accelerated relativistic particles.  Insofar as the relativistic particles are fast-cooling and will generate gamma-rays/neutrinos in direct proportion to $E_{\rm rel}$ (the calorimeteric limit; \citealt{Metzger+15}), this in turn implies that the total optical energy of all shock-powered transients in the universe places an upper bound on the gamma-ray/neutrino background given some assumption about the value of $\epsilon_{\rm rel}$ and the spectrum of non-thermal particles (in our case motivated by observations of novae).  This is the main technique applied in this paper.  

Before proceeding, we must prove several assumptions made above, using $t \sim t_{\rm pk}$ (eq.~\ref{eq:tpk}) as the critical epoch at which we must check their validity.  Firstly, consider the assumption that the shocks are radiative.  Thermal gas behind the shock will cool radiatively on a timescale
\begin{equation}
    t_{\rm cool}=\frac{\mu}{\mu_p}{\mu_e}\frac{3\,k\,T_{\rm sh}}{8\Lambda n_{\rm sh}} = \frac{9}{128}\frac{m_p v_{\rm sh}^{2}}{\Lambda n_{\rm sh}}
\label{eq:tcooldef}
\end{equation}
where $\Lambda$ is the cooling function at $T = T_{\rm sh}$ and we have evaluated $T_{\rm sh}$ using equation (\ref{eq:Tsh}).  Here $\mu_e = 2/(1+X) \simeq 1.16$ and $\mu_p = 1/X \simeq 1.39$ for hydrogen mass fraction $X = 0.72$.  At high temperatures $T \gtrsim 10^{7.3}$ K free-free cooling dominates, for which $\Lambda \approx \Lambda_{\rm ff}\approx 2.3\times10^{-27}\,(T_{\rm sh}/$K$)^{1/2}\,\rm erg\,cm^3\,s^{-1}$  \citep{Draine11}.\footnote{At lower temperatures, $10^{5} < T < 10^{7.3}$ K, cooling from line emission also contributes, with $\Lambda_{\rm line} \approx 1.1\times 10^{-22}(T_{\rm sh}/{\rm K})^{-0.7}$ erg\,cm$^3$\,s$^{-1}$ \citep{Draine11}.}   The ratio of cooling to the shock dynamical timescale is thus
\begin{eqnarray}
\left.\frac{t_{\rm cool}}{t_{\rm dyn}}\right|_{t_{\rm pk}} &=& \frac{9}{128}\frac{\kappa_{\rm opt}}{c}\frac{m_p^{2} v_{\rm sh}^{4}}{\Lambda} \underset{\Lambda \approx \Lambda_{\rm ff}}\approx 10^{-3}\kappa_{\rm 0.3}v_{8.5}^{3},
\label{eq:tcool}
\end{eqnarray}
where we have normalized $\kappa_{\rm opt} = 0.3\kappa_{\rm 0.3}$ cm$^{2}$ g$^{-1}$ to a characteristic optical opacity similar to the electron scattering value for fully ionized gas $\kappa_{\rm es} \simeq \sigma_{\rm T}/m_p \simeq 0.38$ cm$^{2}$ g$^{-1}$, a reasonable approximation for hydrogen-rich ejecta; however, the opacity may be somewhat lower due to lower ionization in the case of hydrogen-poor supernovae (e.g., SLSNe-I) where it may instead result from Doppler-broadened Fe lines (e.g.~\citealt{Pinto&Eastman00}).  From equation (\ref{eq:tcool}) we conclude that the shocks are generically radiative ($t_{\rm cool} \ll t_{\rm dyn}$) at the epoch of peak light/relativistic-particle acceleration, for shock velocities $v_{\rm sh} \lesssim$ 10,000-30,000 km s$^{-1}$, which agrees with the findings of \citep{Murase+11, 2013ApJ...769L...6K, 2014MNRAS.440.2528M}.  

What about the non-thermal particles?  Relativistic ions accelerated at the shock (when it becomes collisionless at times $t \gtrsim t_{\rm pk}$) will carry a power given by $L_{\rm rel} \approx \epsilon_{\rm rel}L_{\rm sh}$ and an total energy $E_{\rm rel}$ (eq.~\ref{eq:Erel}), where $\epsilon_{\rm rel} \sim 0.003-0.01$ in novae (\S\ref{sec:novae}).  After escaping the shock upstream into the unshocked ejecta, or being advected downstream into the cold shell, the relativistic ions will undergo inelastic collisions with ambient ions, producing pions and their associated gamma-ray and neutrino emission.\footnote{Photohadronic interactions with the supernova optical light can be shown to be highly subdominant compared to p-p interactions.}  This interaction occurs on a timescale, $t_{\rm pp} \approx (n\, \sigma_{\rm pp}\, c)^{-1},$ where $\sigma_{\rm pp}\approx 5\times10^{-26}\,\rm cm^2$ is the inelastic proton-proton cross section around 1~PeV \citep{PDG}.  Again, considering the ratio   
\begin{eqnarray}
\left.\frac{t_{\rm pp}}{t_{\rm dyn}}\right|_{t_{\rm pk}} = \left(\frac{v_{\rm sh}}{c}\right)^{2}\left(\frac{\sigma_{\rm opt}}{\sigma_{\rm pp}}\right)\approx   10^{-3}\kappa_{\rm 0.3}v_{8.5}^{2},
\label{eq:tpp}
\end{eqnarray}
 we see that $t_{\rm pp} \ll t_{\rm dyn}$ for $v_{\rm sh} \lesssim 30,000$ km s$^{-1}$.  As in the case of thermal particles, relativistic particles (above the threshold energy) will pion produce on a timescale much shorter than they would lose their acquired energy to adiabatic expansion of the ejecta.\footnote{In principle, energetic particles near the maximum energy (see eqs.~\ref{eq:Emax1}, \ref{eq:Emax2}) could freely stream away from the shock at the speed of light rather than being trapped and advected towards the central shell, in which case they could in principle escape the medium without pion production.  However, this escaping fraction is likely to be small at energies $\lesssim E_{\rm max}$ and account for a small fraction of the total energy placed into relativistic particles \citep{Metzger+16}.}

Protons may also interact with the ambient photons through photopion production when their energy is above the pion production threshold, $E_{p,\rm th} \approx (\epsilon_{p\gamma,\rm th}/ \epsilon_{\rm opt}) \,m_p\,c^2 = 1.4 \times 10^{16}\,(\epsilon_{\rm opt}/10\,\rm eV)^{-1}\,\rm eV$, with $ \epsilon_{p\gamma,\rm th} = (m_\pi + m_\pi^2 / m_p)c^2\approx 150\,{\rm MeV}$. When the photopion production is allowed, it may play an important role with a competing timescale comparing to the pp interaction,   
\begin{equation}
\label{eqn:t_pgamma}
\left.\frac{t_{pp}}{t_{p\gamma}}\right |_{t_{\rm pk}} = \frac{9}{32}f_\Omega \frac{m_p v_{\rm sh}^2}{\epsilon_{\rm opt}}\frac{\sigma_{p\gamma}}{\sigma_{pp}} = 4\,v_{\rm sh, 8.5}^2\,\epsilon_{\rm opt, 1}^{-1}f_\Omega,
\end{equation}
where $\sigma_{p\gamma}\approx 70\,\mu b$ is the inelastic photopion interaction cross section \citep{2009herb.book.....D}. 
For most of the parameter space in consideration, the threshold energy can only be reached when $f_\Omega\ll 1$.  We thus do not account for the neutrino production from the photopion production in the calculation below.

The charged pions created by p-p interactions may themselves interact with background protons, at a rate $t_{\pi p}^{-1}\approx (n\,\sigma_{\pi p}\,c)$, or produce Synchrotron radiation, at a rate $t_{\pi, \rm syn}^{-1} = 4\sigma_T u_B c \gamma_\pi (m_e/m_\pi)^2/(3m_\pi c^2)$. In the above expressions  $\sigma_{\pi p}\approx 4\times 10^{-26}\,\rm cm^2$ is the inelastic pion-proton cross section around 0.1-1~PeV \citep{PDG}, $u_B = B^2/(8\pi)$ is the magnetic field energy density, with $B$ defined later in equation~\ref{eq:Bsh}.  However, these interaction timescales,
\begin{eqnarray}
    \left.\frac{t_{\pi p}}{\gamma_\pi\tau_\pi} \right|_{t_{\rm pk}} &=& \left(\frac{v_{\rm sh}}{c}\right)^{2}\left(\frac{\sigma_{\rm opt}}{\sigma_{\rm \pi p}}\right)\left(\frac{t_{\rm pk}}{\gamma_\pi\tau_\pi}\right) \\ \nonumber
    &\approx& 2\times10^{5}\kappa_{\rm 0.3}v_{8.5}^{2}\gamma_{\pi,6}^{-1}t_{\rm pk, month},
\end{eqnarray}
and 
\begin{eqnarray}
     \left.\frac{t_{\pi,\rm syn}}{\gamma_\pi\tau_\pi} \right|_{t_{\rm pk}} = 9\times10^{7}\,\gamma_{\pi, 6}^{-2} \epsilon_{B, -2}^{-1}\kappa_{0.3} t_{\rm pk, month} 
\end{eqnarray}
are much longer than the charged pion lifetime $\gamma_\pi \tau_\pi$, where  $\tau_\pi = 2.6\times10^{-8}\,\rm s$  is the average life time of charged pions at rest and $\gamma_\pi = 10^6\,\gamma_{\pi, 6}$ is a typical Lorentz factor.  Similarly, one can show that around the peak time, muons also quickly decay into neutrinos without much cooling.

 Equations (\ref{eq:tcool}) and (\ref{eq:tpp}) show that both thermal and non-thermal particles cool effectively instantaneously at the epoch of peak shock power, thus forming the theoretical basis for using shock-powered transients as cosmic ray calorimeters \citep{Metzger+15}.  

\subsection{Maximum Ion Energy}

In the paradigm of diffusive shock acceleration, as cosmic rays gain greater and greater energy $E$ they can diffuse back to the shock from a greater downstream distance because of their larger gyroradii $r_{\rm g} = E/(ZeB_{\rm sh})$, where $B_{\rm sh}$ is the strength of the turbulent magnetic field near the shock and $Ze$ is the particle charge.  A promising candidate for generating the former is the hybrid non-resonant cosmic-ray current-driven streaming instability (NRH; \citealt{Bell04}).  The magnetic field strength near the shock may be estimated using equipartion arguments:
 
\begin{eqnarray}\label{eq:Bsh}
    B_{\rm sh} = \left(6\pi\,\epsilon_B\,m_p\,n_{\rm sh}\,v_{\rm sh}^2\right)^{1/2}, 
\end{eqnarray}
where $\epsilon_{B} \ll 1$ is the ratio of the magnetic energy density to the immediate post-shock thermal pressure.

The maximum energy to which particles are accelerated before escaping the cycle, $E_{\rm max}$, is found by equating the upstream diffusion time $t_{\rm diff} \sim D/v_{\rm sh}^{2}$ with the downstream advection time $t_{\rm adv} \sim \Delta R_{\rm acc}/v_{\rm sh}$, where $\Delta R_{\rm acc}$ is the width of the acceleration zone.  Taking $D \approx r_{\rm g} c/3$ as the diffusion coefficient \citep{Caprioli&Spitkovsky14b}, one obtains 
\begin{equation}
E_{\rm max} \sim \frac{3eZ B_{\rm sh}v_{\rm sh}\Delta R_{\rm acc}}{c}
\label{eq:Emax}
\end{equation}

What is the appropriate value of $\Delta R_{\rm acc}$?  In the case of fully-ionized, non-radiative (adiabatic) shocks, it may be justified to take $\Delta R_{\rm acc} \sim R_{\rm sh}$, i.e. to assume that particle acceleration occurs across a large fraction of the system size.  However, in shock-powered transients, the high gas densities result in very short radiative recombination times, rendering the gas far upstream or downstream of the shock quasi-neutral.  Neutral gas is challenged to support a strong magnetic field, and ion-neutral damping can suppress the growth of the NRH \citep{Reville+07}.  Indeed, in novae the temperature ahead of the shocks may in some cases be too low for efficient collisional ionization, in which case the radial extent of $\Delta R_{\rm acc}$ into the upstream flow is a narrow layer ahead of the shock which has been photo-ionized by the shock's UV/X-ray emission \citep{Metzger+16}.  

In luminous extragalactic transients with high effective temperatures near optical peak$-$the main focus of this paper$-$ionization is less of a concern than in novae.  However, the maximal extent of the particle acceleration zone behind the shock is still limited because of thermal cooling, which compresses the length of the post-shock region to a characteristic width $\Delta R_{\rm cool} \sim v_{\rm sh}t_{\rm cool}$, where $t_{\rm cool}$ is defined in equation (\ref{eq:tcooldef}).  Taking $\Delta R_{\rm acc} = \Delta R_{\rm cool}$ in equation (\ref{eq:Emax}) we obtain \footnote{Although the magnetic field behind the shock may increase due to flux
conservation as gas cools and compresses, this is unlikely to result
in an appreciably larger $E_{\rm max}$ than we have estimated because the
ratio of the Larmor radius to the thermal cooling length $\Delta R_{\rm cool}
\propto 1/n\Lambda$ (which controls the radial width of the cooling
region at a given temperature/density) will decrease moving to higher
densities $n \gg n_{\rm sh}$ relative to its value immediately behind
the shock.} 
\begin{eqnarray}
E_{\rm max} &\sim& \frac{3eZ Bv_{\rm sh}R_{\rm sh}}{c}\frac{t_{\rm cool}}{t_{\rm dyn}} \nonumber \\
&\approx& \frac{eZ}{c}\left(\frac{48 \epsilon_B v_{\rm sh}L_{\rm sh}}{f_{\Omega}}\right)^{1/2}\left(\frac{t_{\rm cool}}{t_{\rm dyn}}\right) 
\label{eq:Emax1}
\end{eqnarray}
where in the second line we have used equations (\ref{eq:Lsh}) and (\ref{eq:Bsh}).  Evaluating this at $t = t_{\rm pk}$, we find
\begin{eqnarray}
E_{\rm max}|_{t_{\rm pk}} \approx  3\times 10^{14}{\rm eV}\,\,Z&\epsilon_{B,-2}^{1/2}f_\Omega^{-1/2}\kappa_{0.3}L_{\rm sh,43}^{1/2}v_{8.5}^{7/2},
\label{eq:Emax2}
\end{eqnarray}
where $\epsilon_{B,-2} \equiv \epsilon_B/(10^{-2}$), $L_{\rm sh,43} = L_{\rm sh}/(10^{43}$ erg s$^{-1}$), and we have used equation (\ref{eq:tcool}) for $t_{\rm cool}/t_{\rm dyn}$.

For a large shock velocity, the proton-proton interaction time may be shorter than the advection time across the cooling length $t_{\rm pp}<t_{\rm cool}$. In this regime, the maximum energy is determined by  $t_{\rm diff} \sim t_{\rm pp}$ and we can obtain a similar form as in equation~\ref{eq:Emax1}, 
\begin{equation}
    E_{\rm max} \sim \frac{3eZ Bv_{\rm sh}R_{\rm sh}}{c}\frac{t_{\rm pp}}{t_{\rm dyn}},
\end{equation}
with $t_{\rm pp}/t_{\rm dyn}$ at the peak time evaluated in equation~\ref{eq:tpp}.

Thus, $E_{\rm max}$ is a very sensitive function of the shock velocity.  Since in most cases $v_{\rm sh}$ and $L_{\rm sh}$ will decrease as the shock sweeps up gas (and since non-thermal particle acceleration cannot occur at times $t \ll t_{\rm pk}$), then $E_{\rm max}|t_{\rm pk}$ is a reasonably good proxy for the maximum particle energy achieved over the entire shock interaction.

The inelastic collisions of ions of energy $E$ with ambient ions to generate $\pi^{0}$($\pi^{\pm}$) will typically produce gamma-rays(neutrinos) of energy $\sim 0.1\,E$~($0.05\,E$) \citep{Kelner&Aharonian08}.  Given the characteristic values up to $E_{\rm max} \gtrsim 10^{16}$ eV implied by equation (\ref{eq:Emax2}) for characteristic velocities $\bar{v}_{\rm ej} \sim v_{\rm sh} \gtrsim 10,000$ km s$^{-1}$ and luminosities $L_{\rm sh} \sim L_{\rm pk} \sim 10^{44}$ erg s$^{-1}$ of the most luminous astrophysical transients (e.g.~TDEs and SLSNe) under the assumption their light curves are shock-powered, we see that high-energy photons and neutrinos ranging in energy from $\sim 1$ GeV to $\gtrsim 1$ PeV can plausibly be produced. Equation (\ref{eq:Emax2}) also suggests that past energetic supernovae in the Galaxy can contribute to cosmic rays around the knee  \citep{2003A&A...409..799S, 2014MNRAS.440.2528M}, an energy range that can hardly be reached by supernova remnants \citep{2013MNRAS.431..415B}. 

Unfortunately, the covering fraction of the shocks $f_{\Omega}$ entering equation (\ref{eq:vsh}) cannot be directly inferred from observations in most cases.  To evaluate the uncertainty in its value we consider two limits: (1) spherically symmetric interaction (maximal $f_{\Omega} = 1$), which for some transients will result in a value of $v_{\rm sh}$ estimated from equation (\ref{eq:vsh}) which is smaller than the average expansion velocity of the ejecta as measured by optical spectroscopy, $\bar{v}_{\rm ej}$; (2) A covering fraction $f_{\Omega} \le 1$ chosen such that $v_{\rm sh} \le \bar{v}_{\rm ej}/2$, which is the smallest allowed value consistent with some characteristic ejecta speed $\bar{v}_{\rm ej}$ (since the shock cannot be moving faster than the ejecta accelerating it).  In most cases, $\bar{v}_{\rm ej}$ should be taken to be the kinetic-energy weighted average velocity; although the ejecta may contain a tail of much faster ejecta (or which covers a very limited solid angle $f_{\Omega} \ll 1$, e.g.~a collimated jet), such shocks may not dominate the total energetics and hence are less relevant to our analysis.  These limits define an uncertainty range of $v_{\rm sh}$ which from equation (\ref{eq:Emax2}) in turn translates into a range of $E_{\rm max}$.

\subsection{Gamma-Ray Escape}\label{subsec:gammaEscape}

Although neutrinos readily escape the ejecta without being absorbed, gamma-rays may have a harder time.  

For relatively low-energy gamma-rays, the dominant source of opacity is Compton scattering off electrons in the ejecta, for which the cross-section in the Klein-Nishina regime ($x \equiv E_{\gamma}/m_e c^{2} \gg 1$, where $E_{\gamma}$ is the gamma-ray energy) is approximately given by $\sigma_{\rm KN} = (3/8)(\sigma_{\rm T}/x)({\rm ln}[2x] + 1/2)$.  Given that $\tau_{\rm T} \lesssim few$ at the epoch of peak optical and gamma-ray emission, attenuation by Compton scattering is generally not important at the gamma-ray energies $E_{\gamma} \gtrsim 100$ MeV of interest.

Gamma-rays can also interact with the nuclei in the ejecta through the Bethe-Heitler (BH) process, for which the cross section can be approximated as \citep{1992ApJ...400..181C}:
\begin{equation}
    \sigma_{\rm BH} \simeq \frac{3}{8\pi}\alpha \sigma_T\,Z^2\, \left[\frac{28}{9}\,\ln(2\,x)-\frac{218}{27}\right]
\end{equation}
where $\alpha \simeq 1/137$ and $Z$ is the atomic charge of the nuclei of atomic weight $A$ (not to be confused with the wind-loss parameter).   Using condition (\ref{eq:tpk}), the BH optical depth $\tau_{\rm BH} \equiv \Sigma \sigma_{\rm BH}/A$ near peak light at photon energies $x \gg 1$ can be written as,
\begin{eqnarray}
\tau_{\rm BH}|_{\rm t_{\rm pk}} &\approx& \left(\frac{c}{v_{\rm sh}}\right)\left(\frac{\sigma_{\rm BH}/A }{\sigma_{\rm opt}}\right) \\ \nonumber
&\approx& 0.3\frac{Z_{\rm eff}^{2}}{A_{\rm eff} }v_{8.5}^{-1}\kappa_{0.3}^{-1}f_{\rm BH},
\label{eq:tauBH}
\end{eqnarray}
where $f_{\rm BH}(x) \equiv \left[\ln(2\,x)-\frac{109}{42}\right]$ and $Z_{\rm eff}/A_{\rm eff}$ are average effective atomic charge/mass of the ejecta ($A_{\rm eff} = Z_{\rm eff} \approx 1$ for H-rich SNe; $A_{\rm eff} = 2Z_{\rm eff} \approx 16$ for the oxygen-rich ejecta of stripped-envelope SNe).

Thus, depending on the shock velocity we see that$-$at the epoch of peak light and particle acceleration$-$we can have $\tau_{\rm BH} \gtrsim 1$ at photon energies $\gtrsim$ GeV ($x \gtrsim 10^{3}$), especially for hydrogen-poor explosions with lower opacity $\kappa \lesssim 0.03$ and metal-rich ejecta with high $Z$.

Gamma-ray photons can also be attenuated due to $\gamma-\gamma$ pair production with ambient photons (e.g.~\citealt{Cristofari+20}).  The optical depth for interaction on the reprocessed optical light from the transient near peak light can be written $\tau_{\gamma\gamma} \sim \sigma_{\gamma\gamma} \,n_{\rm opt}\,R_{\rm sh}$, where $n_{\rm opt} \approx L_{\rm sh}\tau_{\rm pk}/(4\pi R_{\rm sh}^{2} c \varepsilon_{\rm opt}$) is the radiation density, $L_{\rm sh} \approx L_{\rm pk}$ is the optical luminosity assuming it to be shock-dominated, $\tau_{\rm pk} \approx c/\bar{v}_{\rm ej}$, $\varepsilon_{\rm opt} \approx 3kT_{\rm pk} = \varepsilon_{\rm opt, 1}\, 10$ eV is the characteristic energy of a UV/optical photon near the shock (where $T_{\rm pk} \approx \tau_{\rm pk}^{1/4}[L_{\rm sh,pk}/(4\pi \sigma v_{\rm sh}^{2}t_{\rm pk}^{2})]^{1/4}$), and $\sigma_{\gamma\gamma}\approx (3/16)\,\sigma_T$ is the cross section near the pair-production threshold, which occurs for particle energies $E_{\rm \gamma\gamma,th} \approx 2(m_e\,c^2)^2/\varepsilon_{\rm opt} \approx 0.05\,\varepsilon_{\rm opt, 1}^{-1}$ TeV.  Again evaluated around the epoch of peak light and particle acceleration,
\begin{eqnarray}
\tau_{\rm \gamma \gamma}|_{t_{\rm pk}} &\approx& \frac{27}{512}f_\Omega\frac{ m_p v_{\rm sh}^{2}}{\varepsilon_{\rm opt}}\frac{\sigma_{\rm T}}{\sigma_{\rm opt}}\left(\frac{c}{v_{\rm sh}}\right) 
\nonumber \\ \nonumber
&\approx& 7\times10^{4} f_\Omega v_{8.5}\kappa_{0.3}^{-1}\varepsilon_{\rm opt, 1}^{-1}.
\end{eqnarray}
Thus, photons of energy $\gtrsim E_{\rm \gamma\gamma,th}  \sim 0.1-1$ TeV will generally be attenuated before escaping.\footnote{Gamma-rays with lower energies can in principle pair-produce on harder UV/X-rays of energy $\lesssim kT_{\rm sh}$ (eq.~\ref{eq:Tsh}) which exist immediately behind the shocks.  However, due to the thin geometric extent of the cooling layer, and the lower number density of high-energy photons carrying the same luminosity, this form of attenuation is sub-dominant compared to other forms of opacity in this energy range (e.g., inelastic Compton scattering; as also noted in \citealt{Murase+11}); see Fig.~\ref{fig:tau_gamma}.  }

\subsection{Example Shock-Powered Transient}
\begin{figure}
\includegraphics[width= \linewidth] {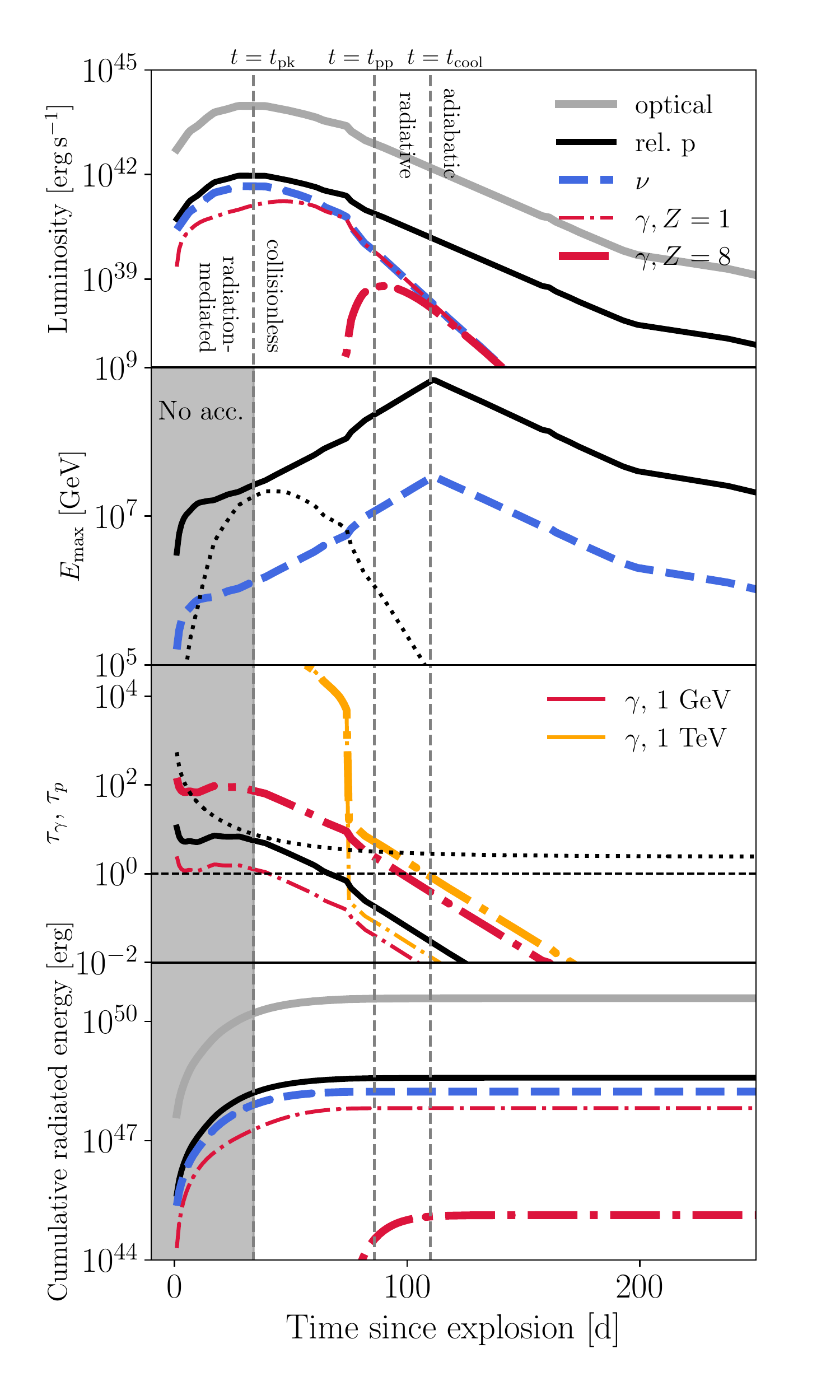} 
\caption{\label{fig:L_t}
Example shock-powered optical transient, showing the dependence of various quantities related to relativistic particle acceleration as a function of time since explosion.  From top to bottom: (1) luminosities of shocks (reprocessed optical emission) and their observable signatures (relativistic protons, neutrinos and $\gamma$-rays, in the latter case for different assumptions about the ejecta composition); (2) maximum accelerated proton energy and emitted neutrino energy; (3) optical depths, of protons to p-p interactions, 1~GeV and 1~TeV $\gamma$-rays (shown separately for H-rich and H-poor shocked medium as denoted by different values of the ejecta composition $Z$); and (4) cumulative radiated energy in the form of optical emission (grey), relativistic protons (black), neutrinos (blue dashed) and $\gamma$-rays (red and orange dash-dotted).   We have adopted a canonical Type II SLSNe light curve from \citet{Inserra19}.  Dotted lines in panel 2 and 3 show how the evolution of $E_{\rm p, max}$ and the p-p interaction optical depth would instead change if the luminosity evolution is driven by a decelerating shock (decreasing $v_{\rm sh}$) into a medium of constant wind parameter $A$.  The true evolution of the shock properties likely lie between these two limits, i.e. $E_{\rm p, max}$ relatively constant in time.  Relativistic particle acceleration, and thus $\gamma$-ray/neutrino emission, is not expected prior to the optical peak (shown as a gray shaded region) due to the shock being radiation-mediated at high optical depths.}
\end{figure}

\begin{figure}
\includegraphics[width= \linewidth] {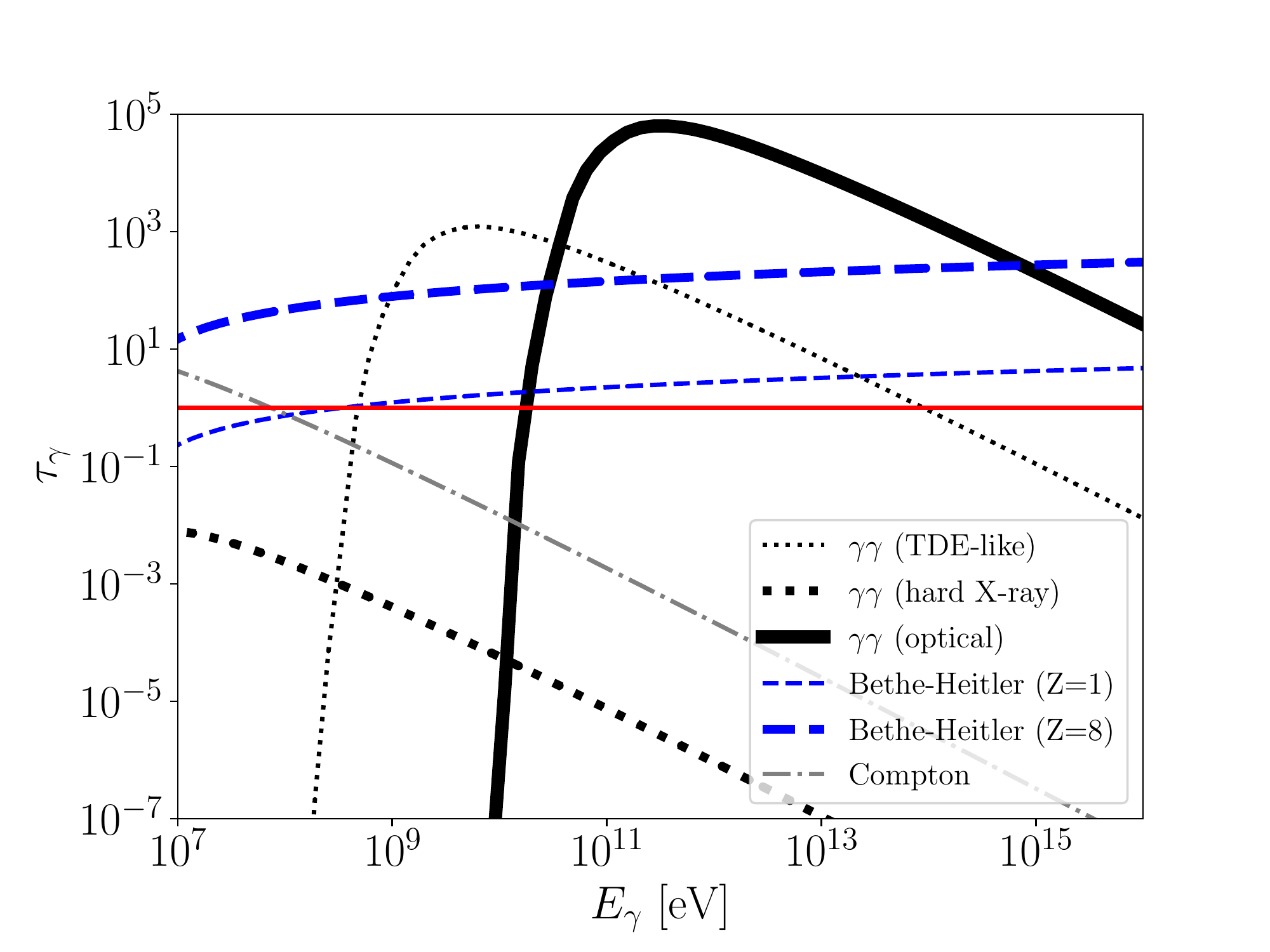} 
\caption{\label{fig:tau_gamma}
 Optical depth of the ejecta to $\gamma$-rays as a function of the gamma-ray energy $E_{\gamma}$, evaluated for conditions corresponding to the the example shock-powered transient in Fig.~\ref{fig:L_t} around the epoch of peak light and particle acceleration, $t \approx t_{\rm pk}$.  Blue dashed lines show the Bethe-Heitler optical depth for two different assumptions about the nuclear composition of the ejecta ($Z = 1, 8$), while a grey dash-dotted line shows the effective optical depth due to Compton scattering.   Solid, large and small} dotted black lines show the optical depth to $\gamma-\gamma$ pair production off of the optical, X-ray and a TDE-like (peaked around 100~eV) thermal radiation, respectively. For comparison, the red solid line indicates $\tau_\gamma=1$. 
\end{figure}
As an example of a shock-powered transient, Figure~\ref{fig:L_t} presents the time-evolution of the luminosities (top panel) and cumulative radiated energies (bottom panel) in optical, relativistic protons, neutrinos and $\gamma$-rays.  We consider a SLSN-II event with $L_{\rm pk} = 10^{44}\,\rm erg\,s^{-1}$, $t_{\rm pk} = 34$~d and $\bar{v}_{\rm ej} = 8000\,\rm km\,s^{-1}$ \citep{Inserra19}, with a characteristic optical light curve from \citet{Inserra19}. The optical luminosity, which well represents the shock power after $t_{\rm pk}$, is used to evaluate $\bar{v}_{\rm ej}(t)$ and $A(t)$ using equation~\ref{eq:Lsh}.  To break the degeneracy of the time dependence, we consider two limits, wherein either $\bar{v}_{\rm ej}$ or $A$ is assumed to be constant in time.  Most curves in the figure correspond to the former limit ($\bar{v}_{\rm ej} = const$), except the black dotted curves in the second and third panels (which assume $A = const$).  

The luminosity of relativistic protons, $L_p\equiv L_{\rm rel}$, is computed using equation~\ref{eq:Erel} with $f_{\rm sh} = 1$ and $\epsilon_{\rm rel} = 0.01$ (see \S\ref{sec:novae}).   As proton-proton interactions roughly equally split the proton energy into into neutrinos and electromagnetic energy ($\gamma$-rays and electrons), the neutrino and $\gamma$-ray luminosities are evaluated as 
\begin{equation}\label{eqn:L_nu}
    L_\nu \approx 1/2\,L_p \,f_{\rm pp}
\end{equation}
and 
\begin{equation}
    L_\gamma \approx 1/3\,L_p\,f_{\rm pp} \,\exp({-\tau_\gamma})
\end{equation}
respectively. The factor $1/2$ arises because charged pions are produced with roughly 2/3 probability in a pp interaction and about three quarters of their energy is carried away by neutrinos. The other quarter is carried away by electrons. These electrons, with energy $\approx 50\,(E_p / 1\,\rm PeV)$~TeV, lose most of their energy through Synchrotron radiation, as their inverse Compton process with optical photon background is suppressed due to the Klein-Nishina effect.
The factor $1/3$ in $\gamma$-ray spectrum is because neutral pions are produced with roughly 1/3 chance and all their energy is carried by photons.
The maximum proton energy, $E_{\rm max}$, is computed from equation~\ref{eq:Emax1} for $\epsilon_{\rm B} = 0.01$, and the radiated neutrino energy is estimated as $E_\nu \approx 0.05\,E_p$. $f_{\rm pp} = 1-\exp(-\tau_{\rm pp})$ is the pion production efficiency at $E_p\sim E_{\rm max}$, where $\tau_{\rm pp} \approx n_{\rm sh}\,\sigma_{\rm pp}\,R_{\rm sh}$ and $\tau_\gamma$ are the optical depth of relativistic protons and $\gamma$-rays, respectively.
At lower energy, $E_p\ll E_{\rm max}$, protons are trapped and advected at the shock velocity, so the pion production efficiency at these energies is instead $f_{\rm pp}=1-\exp(-t_{\rm dyn}/t_{\rm pp})=1-\exp(-\tau_{\rm pp}c/v_{\rm sh})$. The correction to $f_{\rm pp}$ barely affects the neutrino flux calculation since $\tau_{\rm pp} > 1$ around the peak time when most neutrinos are produced. It may however significantly increase the $\gamma$-ray flux in a scenario where most $\gamma$-rays are produced at late time. 

Figure \ref{fig:tau_gamma} show the optical depth of the ejecta as a function of gamma-ray energy $E_{\gamma}$ at an epoch around optical peak ($t \approx t_{\rm pk}$) for each of the processes described above.  The third panel in Figure~\ref{fig:L_t} show the optical depth of the ejecta to gamma-rays of energy $E_{\gamma} = 1$ TeV and $E_{\gamma} = 1$ GeV, the latter for two different choices of the nuclear composition of the ejecta, $Z = 1, 8$ (corresponding roughly to hydrogen-rich and hydrogen-poor explosions, respectively).  Due to the bright optical background, TeV $\gamma$-rays are heavily attenuated by pair production in the first $\sim 90$ days. After that optical photons fall below the energy threshold needed for pair production with TeV photons.  The attenuation of GeV $\gamma$-rays is dominated by the Bethe-Heitler process.  Depending on the composition of the external medium, the source is $\gamma$-ray dark in the first $\sim$50 to $\sim$100 days.  As a result, although the total radiated energy in neutrinos is a fixed fraction $\sim \epsilon_{\rm rel}/2$ of the total optical output and saturates quickly around $t_{\rm pk}$ (bottom panel of Fig.~\ref{fig:L_t}), the total radiated energy in gamma-rays is greatly suppressed, particularly in the case of hydrogen-poor external medium ($Z_{\rm eff} = 8$).

\section{Particle Acceleration in Novae}
\label{sec:novae}

Classical novae observed simultaneously via their optical and high-energy gamma-ray emission offer an excellent opportunity to test and calibrate our understanding of particle acceleration at internal radiative shocks.  The brightest novae achieve peak optical luminosities $L_{\rm pk} \sim 10^{38}-10^{39}$ erg s$^{-1}$ and light curves that rise on a timescale $t_{\rm pk} \sim$ days $\sim 10^{5}$ s \citep{Gallagher&Starrfield78}.  The tight temporal correlation between the optical and gamma-ray luminosities \citep{Li+17,Aydi+20} strongly suggest that much of the optical luminosity is powered by internal radiative shocks \citep{Metzger+14}, i.e.~$L_{\rm pk} \approx L_{\rm sh}(t_{\rm pk})$.  Using equation (\ref{eq:tpk}) and (\ref{eq:vsh}) with a characteristic covering fraction of the external medium $f_{\Omega} = 0.2$ (e.g.~\citealt{Chomiuk+14,Derdzinski+17}) and $\kappa_{\rm opt} = 0.3$ cm$^{2}$ g$^{-1}$, we derive a value $v_{\rm sh} \sim$ 500 km s$^{-1}$, which is reasonable from optical spectroscopy.  We also find $A \equiv \dot{M}/(4\pi v_w) \approx ct_{\rm pk}/\kappa_{\rm opt} \sim 10^{6}A_{\star}$; taking $v_w \sim v_{\rm sh}$, the latter corresponds to a mass-loss rate $\dot{M} \sim 10^{25}$ g s$^{-1}$ and hence a total mass ejection $\dot{M}t_{\rm pk} \sim 10^{-4}-10^{-3}M_{\odot}$, broadly consistent with that inferred by nova modeling \citep{Gehrz+98}.    

In detail, the simplified set-up laid out in $\S\ref{sec:overview}$ for explosive transients is not wholly applicable to novae because much of the total radiated shock energy occurs after some delay with respect to the optical rise time $t_{\rm pk}$.  Shock interaction in novae is in most cases likely driven by a fast wind from the white dwarf which is observed to {\it accelerate} in time, resulting in higher ejecta speeds and shock velocities $v_{\rm sh} \gtrsim 10^{3}$ km s$^{-1}$ being reached on the timescale of $\sim$ weeks $\gg t_{\rm pk}$ over which most of the gamma-ray emission occurs \citep{Ackermann+14}.  This kind of wind-powered transient behavior is distinct from singular explosive transients like supernovae, for which in general there is no sustained long-lived activity from a ``central engine", such that $v_{\rm sh}$ (and hence $L_{\rm sh}$ for most external medium density profiles) only declines at times $t \gtrsim t_{\rm pk}$.\footnote{This unusual time evolution of the shock power in novae also explains why it is possible for $\sim$ GeV gamma-rays to evade the constraints set by BH absorption (eq.~\ref{eq:tauBH}) and escape from the ejecta.  However, the delayed onset of gamma-ray emission relative to the optical peak seen in some novae (the earliest gamma-ray data in ASASSN16ma provides a striking example; \citealt{Li+17}) may point to absorption occurring around $\sim t_{\rm pk}$ even in these systems.} 

\begin{figure}
\includegraphics[width= \linewidth] {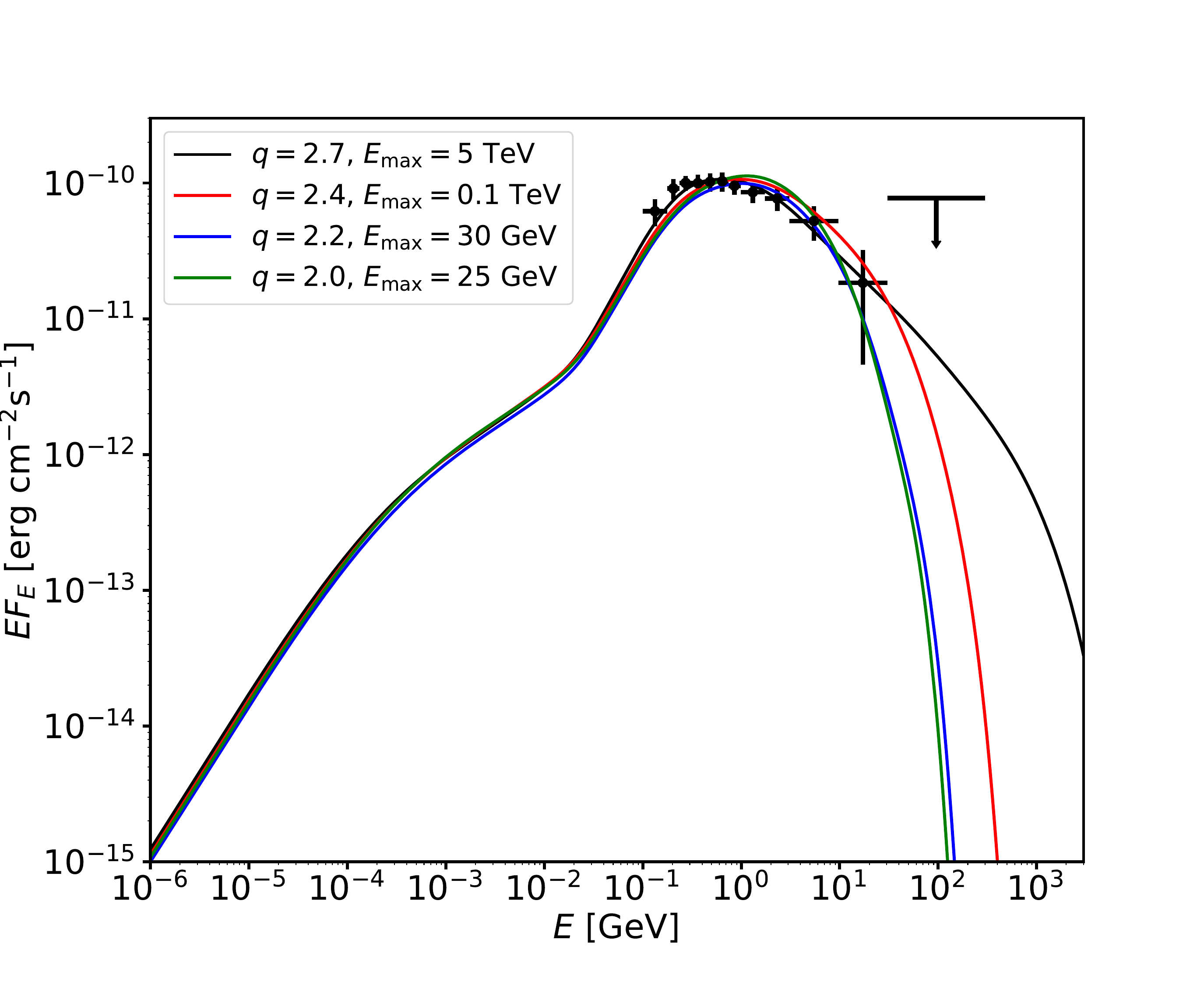} 
\includegraphics[width= \linewidth] {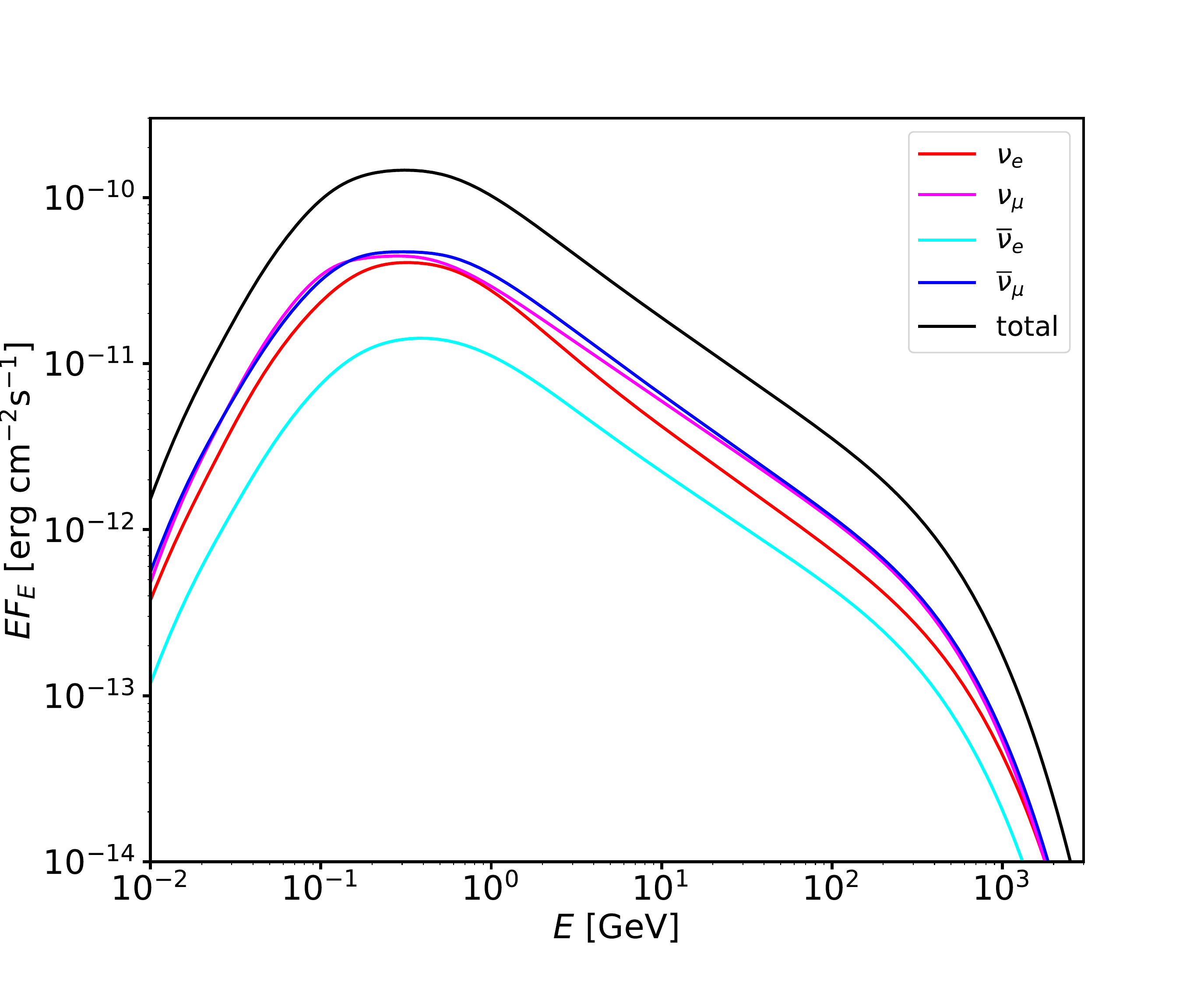} 
\caption{\label{fig:Indrek}
{\bf Top Panel:} Models of hadronic gamma-ray emission from non-relativistic radiative shocks \citep{Vurm&Metzger18} fit to the time-integrated {\it Fermi} LAT spectra of the classical novae ASASSN 16ma (\citealt{Li+17}; black points).  The models make different assumptions about the injected population of relativistic protons at the shock, such as the power-law index $q$ of their momentum spectrum and the high energy cut-off, $E_{\rm max}$.  For low values of $q \approx 2-2.2$ (with $E_p^{2}(dN_p/dE_p) \sim const$) the data require a modest $E_{\rm max} \lesssim 30$ GeV, while for larger $q \gtrsim 2.4$ the value of $E_{\rm max}$ is essentially unconstrained (we take $E_{\rm max} = 5$ TeV in the $q = 2.7$ model).  {\bf Bottom Panel:} Neutrino spectra for the $E_{\rm max} = 5$ TeV, $q = 2.7$ model shown in the top panel.}
\end{figure}

Nevertheless, insofar as we have good evidence that the gamma-ray emission from novae is powered by internal radiative shocks in the calorimetric limit \citep{Metzger+15}, we can use the properties of the particle acceleration as inferred from their observed gamma-ray luminosity and energy spectrum to guide our expectations for shock-powered transients more generally.  
Figure \ref{fig:Indrek} shows models of hadronic gamma-ray emission from radiative shocks calculated based on the models of \citet{Vurm&Metzger18} and applied to the time-integrated gamma-ray spectrum of the nova ASASSN16ma \citep{Li+17}.  The model assumes that protons are injected at the shock with a number distribution $dN_p/dp \propto p^{-q}$, where $p = \beta\gamma m_p c^{2}$ is the proton momentum and $q$ is a power-law index.  The normalization of the accelerated proton energy, $E_{\rm rel}$, is assumed to be proportional to the radiated optical fluence according to $\epsilon_{\rm rel} = E_{\rm rel}/E_{\rm opt}$.  Some models also include an exponential cut-off above the momentum $p_{\rm max} = E_{\rm max}/c$ corresponding to some maximum proton energy, $E_{\rm max}$.

As shown in Figure \ref{fig:Indrek}, several of the models can in principle reproduce the main features of the observed spectrum, particularly the overall spectral shape, including the deficit in the lowest energy bin $\lesssim$~few 100 MeV.  This low-energy turnover arises naturally in hadronic models due to the pion creation threshold corresponding to their rest energy $\sim 135$~MeV; the spectrum in the LAT range is produced mainly by $\pi_0$ decay which generates few photons below this energy. The decay of charged pions $\pi_\pm$ also generates electron-positron pairs of comparable numbers and energies; those contribute mainly in the hard X-ray and MeV domain by inverse Compton and bremsstrahlung, partially suppressed by Coulomb losses.

Although some fits are formally better than others, these differences should not be taken too seriously considering the many simplifications going into the analysis, such as fitting a single set of shock conditions to observations which have been time-averaged over several weeks ($\approx$ many cooling timescales in which the shock properties are likely to evolve).  In all cases we find $\epsilon_{\rm rel} \approx (2-4)\times 10^{-3}$, consistent with the expected acceleration efficiency from corrugated quasi-parallel radiative shocks \citep{Steinberg&Metzger18}.  This is also consistent with upper limits from the Type IIn interacting SN 2010j from {\it Fermi} LAT, which \citet{Murase+19} use to constrain $\epsilon_{\rm rel} \lesssim 0.05-0.1$.  

Figure \ref{fig:Indrek} shows that there exists a significant degeneracy between the value of $q$ and the high-energy cut-off $E_{\rm max}$.  Models with flatter injection (low $q$) require a high-energy cut-off, while for those with steep injection (high $q$) the value of $E_{\rm max}$ is essentially unconstrained.  For instance, both the combinations ($q = 2.4$, $E_{\rm max} = \infty$) and ($q = 2$, $E_{\rm max} \approx 25$ GeV) can fit the data (again, within uncertainties accounting for the simplifying assumptions of the model).  

Despite the above-mentioned degeneracy, there exist theoretical reasons to favor the low $q$ intrinsic cut-off (low $E_{\rm max}$) cases.  Firstly, for high Mach number shocks ($\mathcal{M} \gtrsim 30-100$ in novae) diffusive shock acceleration predicts a spectrum $q \simeq 2$ (e.g.~\citealt{Blandford&Ostriker78,Caprioli&Spitkovsky14}).  Although the spectrum can be steepened by non-linear effects due to cosmic ray feedback on the upstream (e.g.~\citealt{Malkov97}), this is unlikely to be important given the low $\epsilon_{\rm rel} \lesssim 1\%$.  Applying equation (\ref{eq:Emax2}) we find values of $E_{\rm max} \sim 1-100$ GeV for characteristic parameters $L_{\rm pk} \approx 10^{38}-10^{39}$ erg s$^{-1}$, $v_{\rm sh} \approx 500-2000$ km s$^{-1}$, $\kappa_{0.3} \sim 1$, $Z \simeq 1$, $\epsilon_B = 0.01$, consistent with the low $E_{\rm max}$ models in Fig.~\ref{fig:Indrek}.  In principle the high-energy cut-off in nova gamma-ray spectra may not be intrinsic, but instead arise due to $\gamma$-$\gamma$ pair creation on the nova optical light \citep{Metzger+16}; however, this environmental cut-off should not set in until $E_{\gamma} \gtrsim 30$ GeV (Fig.~\ref{fig:tau_gamma}), corresponding to an equivalent $E_{\rm max} \approx 300$ GeV typically higher than needed to fit the data in Fig.~\ref{fig:Indrek}.

Even if proton acceleration in nova shocks ``fizzles out" at $E_{\rm max} \lesssim 100$ GeV, otherwise similar shocks, but scaled to the much higher luminosities needed to power energetic extragalactic transients, could reach significantly higher $E_{\rm max} \propto L_{\rm pk}^{1/2}$ with a flat spectrum $q \simeq 2$.  Motivated thus, in the sections to follow we apply the assumption of moderate $q \lesssim 2.2$ and $E_{\rm max}$ following equation (\ref{eq:Emax2}; for the same value of $\epsilon_{\rm B} = 0.01$ ``calibrated" to match the gamma-ray emission from novae) to extragalactic transients.

\section{Applying the Calorimetric Technique to the Transient Zoo}
\label{sec:zoo}

In this section we apply the basic methodology of \S\ref{sec:overview} to a large range of possible shock-powered transients (several already mentioned in the Introduction) in order to place an upper limit on their high-energy gamma-ray and neutrino emissions.  We do this using exclusively observed properties of each class under the assumption that 100\% of their optical fluence is shock-powered and the particle acceleration properties follow those measured from classical novae.

\subsection{Observed Properties of Transient Classes}

\begin{figure*}
\includegraphics[width= \linewidth] {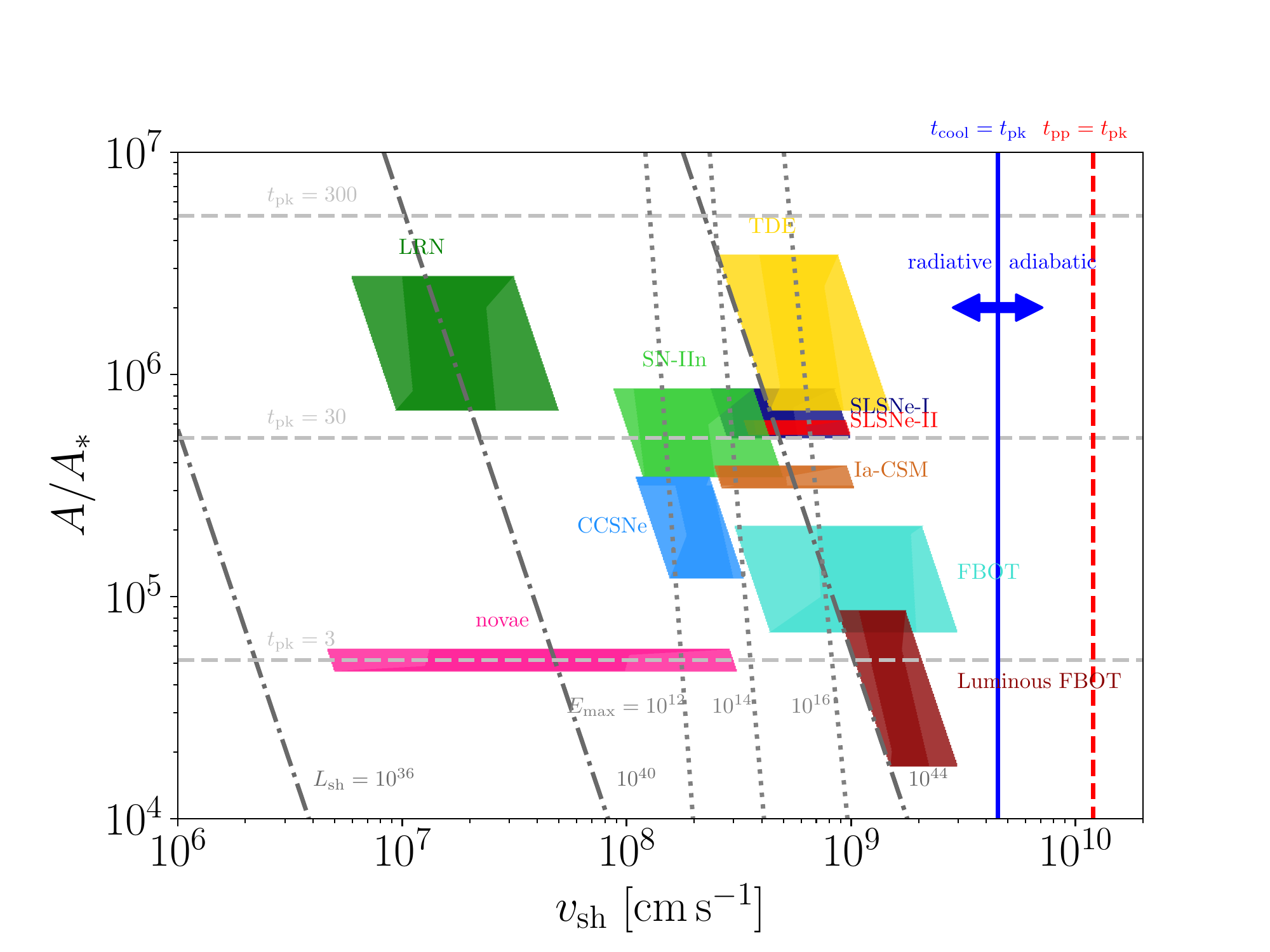} 
\caption{\label{fig:A_v_sh}
Various shock properties in the space of shock velocity $v_{\rm sh}$ and effective wind mass-loss parameter of the external medium $A \equiv \dot{M}/(4\pi f_{\Omega} v_{\rm w})$ normalized to a fiducial value $A_{\star} \equiv 5\times 10^{11}$ g cm$^{-1}$ corresponding to $\dot{M} = 10^{-5}M_{\odot}$ yr$^{-1}$, $v_{\rm w} = 1000$ km s$^{-1}$, $f_{\Omega} = 1$ \citep{Chevalier&Li00}.  Contours show the values of shock luminosity $L_{\rm sh}$ in erg s$^{-1}$, peak time $t_{\rm pk}$ in days and maximum proton energy $E_{\rm max}$ in eV (assuming $\epsilon_B = 0.01$ and $f_\Omega=1$). Color boxes mark the range covered by transients listed in Table~\ref{tab:classes} with $f_\Omega$ in Table~\ref{tab:classes_derived} assuming their light curves are shock-powered (i.e. $L_{\rm pk} = L_{\rm sh}$). The blue and red vertical lines indicate $t_{\rm cool} = t_{\rm pk}$ and $t_{\rm pp} = t_{\rm pk}$, respectively. For typical parameters and at the peak time, all considered transients are in the radiative shock regime and have the hadronuclear interaction time shorter than the dynamical time on the timescale $t_{\rm pk}$ defining the bulk of the thermal and non-thermal radiated energies.
}
\end{figure*}

Table \ref{tab:classes} {\bf and Figure~\ref{fig:A_v_sh} summarize} a diverse list of known or suspected non-relativistic shock-powered optical transients.  For each class, we provide the range of measured or assumed quantities, including the local volumetric rate $\mathcal{R}_0$, peak luminosity $L_{\rm pk}$, peak timescale $t_{\rm pk}$, (kinetic-energy weighted) ejecta velocity $\bar{v}_{\rm ej}$, radiated optical energy $E_{\rm opt}$ (in many cases approximated as $\sim L_{\rm pk}t_{\rm pk}$), and average charge of nuclei $Z_{\rm eff}$ in the ejecta/external medium.  In the final column we also provide a qualitative indicator of our confidence that shock interaction (possibly hidden) plays an important role in powering a sizable fraction of each transient class.  Before proceeding, we go into some details on the various transient classes entering this table.  We also discuss how we expect the rate to evolve with cosmic redshift $z$, as this will enter our background calculations below.  Our main goal is to quantify the total production rate of optical light from different transient classes in order to place constraints on the neutrino background.  

For LRN from stellar mergers, \citet{Kochanek14} find a peak luminosity function $L_{\rm pk}(dN/dL_{\rm pk}) \propto L_{\rm pk}^{-0.4\pm 0.3}$.  Coupled with the tendency for the more luminous LRN to last longer \citep{MetzgerPejcha17}, this suggests a roughly flat distribution of radiated optical energy, i.e. $E_{\rm opt}(dN/dE_{\rm opt}) \sim const$.  As an example to nail the normalization, consider V838 Mon \citep{Munari+02,Tylenda+05}, which peaked at a luminosity $L_{\rm pk} \sim 4\times 10^{39}$ erg s$^{-1}$ on a timescale $t_{\rm pk} \sim 40$ days, corresponding to a total optical output $E_{\rm opt} \sim 10^{46}$ erg. \citet{Kochanek14} estimate a rate of V838 Mon-like transients of 0.03 yr$^{-1}$ in the Milky Way.  Taking a volumetric density of $L_{\star}$ galaxies in the local universe of $\approx$ 0.006 Mpc$^{-3}$, we estimate the local rate of V838 Mon-like LRN of $\mathcal{R}(z=0) \sim 2\times 10^{5}$ Gpc$^{-3}$ yr$^{-1}$. A more detailed analysis would include an integration of the rates over the
distribution of galaxy masses and star formation rates, but given the
significant uncertainty already present in the per-galaxy rate we
neglect this complication here. Since the progenitor of V838 Mon was a relatively massive star binary $\sim 5-10M_{\odot}$ with a short lifetime, the LRN rate will roughly trace the star formation rate (SFR) with redshift. 

For classical novae, the estimated Milky Way rate is $\sim 20-70$ yr$^{-1}$ \citep{Shafter17}.  Again using the $z=0$ density of $L_{\star}$ galaxies, we find a volumetric nova rate of $\sim (1-5)\times 10^{8}$ Gpc$^{-3}$ yr$^{-1}$.  Likewise, at least in irregular and spiral galaxies (which make up an order-unity fraction of stellar mass in the universe), the rate of novae are believed to trace star formation (e.g.~\citealt{Yungelson+97,Chen+16}); hence, to zeroth order novae should also trace the cosmic SFR.  

For TDE flares, \citet{vanVelzen18} find a peak luminosity function $L_{\rm pk}(dN/dL_{\rm pk}) \propto L_{\rm pk}^{-1.5}$ which is dominated by the lowest luminosity events.  The total TDE rate is uncertain, but a value $\sim 10^{-4}$ yr$^{-1}$ per $L_{\star}$ galaxy is consistent with observations \citep{vanVelzen18} and theory (\citealt{StoneMetzger16}; however, the observed preference for post-starburst galaxies is not understood; \citealt{Arcavi+14,Graur+18,Stone+18}). 

For supernovae, we consider separately all core collapse supernovae (CCSNe), which are dominated by Type II SNe with typical values $L_{\rm pk} \sim 10^{42}$ erg s$^{-1}$ and $t_{\rm pk} \sim 100$ d, corresponding to a total radiated output $E_{\rm opt} \sim 10^{49}$ erg.  The Type IIn SN subclass show clear evidence for shock interaction, but not necessarily always at epochs that allow one to conclude it is dominating the total optical output of the supernova (though more deeply embedded shock interaction could be at work during these events).  Following \citet{Li+11} we take the rate of Type IIn SN to be 8.8\% of the total CCSN rate.  

For SLSNe, roughly defined as SNe with peak absolute g-band magnitude $M_g < -19.8$ \citep{Quimby+18}, we take rates of 10-100 Gpc$^{-3}$ yr$^{-1}$ and 70-300 Gpc$^{-3}$ yr$^{-1}$ for the Type I and II, respectively \citep{Quimby+13,GalYam19,Inserra19}.  We do not distinguish between the ``Slow" and ``Fast" sub-classes of SLSNe-I, despite their potentially different physical origins.  A detailed analysis of the luminosity function of SLSNe remains to be performed; however, from the reported population one roughly infers $dN/dL_{\rm pk} \propto L_{\rm pk}^{-\alpha}$ with $\alpha \sim 1$ and hence we pair the events with the lowest(highest) optical fluence with those of the highest(lowest) rate in calculating the fluence-rate below.  

As the name ``Fast Blue Optical Transients" suggests, FBOTs are rapidly-evolving luminous blue transients which can reach peak luminosities similar to SLSNe.  \citet{Coppejans+20} present a summary discussion of FBOT rates.  For all FBOTs with peak g-magnitude in the range $M_g \lesssim -16.5$ ($L_{\rm pk} \gtrsim 10^{43}$ erg s$^{-1}$), \citet{Drout+14} find a rate at $z < 0.6$ of 4800-8000 Gpc$^{-3}$ yr$^{-1}$.  For the most luminous FBOTs with $M_g < -19$ ($L_{\rm pk} \gtrsim 10^{44}$ erg s$^{-1}$), a class including AT2018cow \citep{Prentice+18}, CSS161010 \citep{Coppejans+20}, and ZTF18abvkwla (the ``Koala"; \citealt{Ho+20}), \citet{Coppejans+20} estimate a rate of $\mathcal{R} \sim 700-1400$ Gpc$^{-3}$ yr$^{-1}$ at $z \lesssim 0.2$.  Several of the luminous FBOTs show clear radio signatures of shock interaction on large radial scales \citep{Margutti+19,Ho+20,Coppejans+20}, the energy source behind the bulk of the optical emission in these events is debated (though \citealt{Margutti+19} present evidence that the optical emission in AT2018cow is powered indirectly by reprocessed X-rays).  The association of FBOTs with star-forming host galaxies \citep{Drout+14} again justifies scaling their rate with the cosmic SFR.

A small subset of Type Ia SN show evidence for shock interaction between the ejecta of the exploding white dwarf with hydrogen-rich circumstellar material (so-called ``Type Ia-CSM"; \citealt{Hamuy+03,Chugai&Yungelson04,Aldering+06,Dilday+12,Bochenek+18}).  These events are estimated to accompany between $\sim 0.1-1\%$ of Type Ia SN, corresponding to a volumetric rate of $\sim 300-3000$ Gpc$^{-3}$ yr$^{-1}$.

In addition to the relatively exotic transients above, we also consider the more speculative possibility that even ordinary core collapse supernovae (e.g., Type IIP, Type Ibc) are shock-powered at some level (e.g.,~\citealt{Sukhbold&Thompson17}).\footnote{As an extreme example, the H-rich supernova iPTF14hls, although identical to an ordinary IIP in terms of its spectroscopic properties, exhibited a light curve that stayed bright over 600 days (as opposed to the $\sim 100$ day plateaus of most IIP) with at least 5 distinct peaks \citep{Arcavi+17}.  Although initially there we no spectroscopic indications of shock interaction, emission features finally appeared at late times, revealing a dense CSM \citep{Andrews&Smith18}.} 
  From their explosion models of stripped-envelope stars, \citet{Ertl+20} find that the $^{56}$Ni production in their models is able to explain at best half of the luminosities of Type Ib/c supernovae, pointing to an additional energy source in these systems (see also \citealt{Woosley+20}).

\subsection{Derived Properties of Transient Classes}\label{subsec:derivedTransientQuantities}

Table~\ref{tab:classes_derived} lists several derived properties for each of the transient classes in Table~\ref{tab:classes}, including the local (redshift $z \approx 0$) injection rate of optical energy, $\dot{\cal E}_{\rm opt}$, and the maximum per-particle energy of shock-accelerated protons, $E_{\rm max}$.  The former is calculated according to
\begin{equation}
    \dot{\cal E}_{\rm opt} = {\cal R} (z=0) \,\int dE_{\rm opt}\, E_{\rm opt} \frac{dN}{dE_{\rm opt}}.
\label{eq:dotEopt}
\end{equation}
For all source classes other than CCSNe, we estimate $\dot{\cal E}_{\rm opt}$ using the upper bound of the local rate and the lower bound of the optical energy in Table \ref{tab:classes}, considering that the luminosity function of most transient classes is either flat or dominated by the low-luminosity events (\citealt{2014ApJ...785...28K, vanVelzen18}, also see references in the table). Since CCSNe consist of multiple types of supernovae with each having its own luminosity function \citep{Li+11, 2014ApJ...792..135T}, we multiply the upper bounds of $\cal R$ and $E_{\rm opt}$ to give an optimistic estimate of $\dot{\cal E}_{\rm opt}$.  

The maximum proton energy, $E_{\rm max}$, is calculated following equation~\ref{eq:Emax1} with $Z = 1$.  Although $E_{\rm max} \propto Z$ and hence could be larger for hydrogen-poor CSM, the energy per nucleon $E_{\rm max}/A$ is roughly independent of $Z \simeq A/2$.  As discussed after equation (\ref{eq:Emax2}), the uncertainty in the shock covering fraction $f_{\Omega}$ results in a corresponding uncertainty in $v_{\rm sh} \le \bar{v}_{\rm ej}/2$ (and hence $E_{\rm max}$).  
A smaller $f_{\Omega}$ requires a larger $v_{\rm sh}$ to generate the same optical luminosity.  For transient classes with a range of peak luminosity and peak time, the higher (lower) bounds of $L_{\rm pk}$ are matched with the lower (higher) bounds of $t_{\rm pk}$ to derive the permitted range of $v_{\rm sh}$, $f_\Omega$ and $E_{\rm max}$.

\subsection{$\dot{\cal E}_{\rm opt}$ and $E_{\rm max}$ Required by Neutrino Observation}

\begin{figure*}[htb!]
\includegraphics[width= \linewidth] {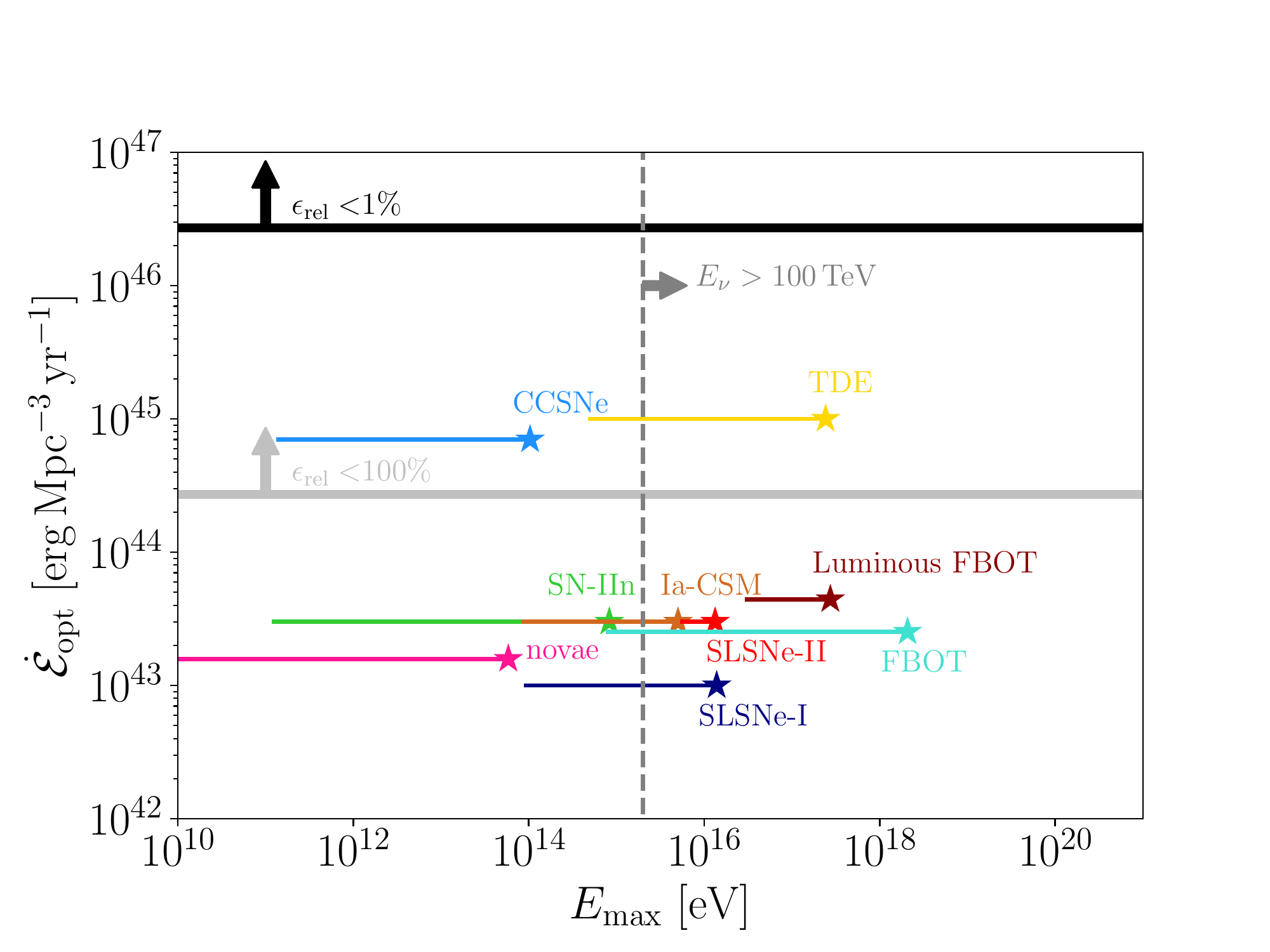} 
\caption{\label{fig:energyrate_Emax}
Injection rate of optical energy, $\dot{\mathcal{E}}_{\rm opt}$ (eq.~\ref{eq:dotEopt}), as a function of maximum accelerated proton energy, $E_{\rm max}$ (eq.~\ref{eq:Emax2}), for various transients with properties in Tables~\ref{tab:classes} and \ref{tab:classes_derived}. 
A range of $E_{\rm max}$ values is shown, encompassing the uncertainty in the covering fraction $f_{\Omega}$ of the shocks (lower $f_{\Omega}$ requires higher velocity shocks$-$leading to larger $E_{\rm max}-$to match the same optical luminosity).   The vertical dashed line indicates the proton energy needed to produce 100~TeV neutrinos.   For comparison, the horizontal lines indicate the energy injection rate required by the IceCube diffuse neutrino background assuming $\epsilon_{\rm rel}=1\%$ (black) and $100\%$ (light gray).  
}
\end{figure*} 

The total neutrino flux contributed by sources over cosmological distances can be calculated by \citet{1999PhRvD..59b3002W}
\begin{equation}
\label{eqn:Phi_nu1}
\Phi(E_\nu)= \frac{{\cal R}_0}{4\pi} \int\,dz\,\frac{c}{(1+z)^2\,H(z)}\, f(z)\,\left(E_\nu^{\prime\,2}\frac{dN}{dE'_\nu}\right)(z),
\end{equation}
where  $E'_\nu = E_\nu (1+z)$ is the redshifted neutrino energy,
$H(z)=H_0\,(\Omega_M\,(1+z)^3+\Omega_\Lambda)^{1/2}$ is the Hubble constant at redshift $z$, ${\cal R}_0$ is the rate of the transient in the local universe, and $f(z)$ describes the source evolution, which equals the source rate at redshift z to that at today, $f(z) = {\cal R}(z)/{\cal R}_0$. As the transient classes in Table~\ref{tab:classes} approximately follow the star formation rate (SFR), we adopt the $f(z)$ from \citealt{2006ApJ...651..142H}.  We adopt a standard cosmology with $H_0=67.4\,\rm km\,s^{-1}\,Mpc^{-1}$ and  $\Omega_m=0.315$ \citep{2018arXiv180706209P}. 

Each transient event provides a total neutrino energy $E_\nu^2({dN}/{dE_\nu}) \approx (1/2)\,f_{\rm pp}\,E_p^2\,{dN}/{dE_p}$, where $dN/dE_\nu$ and $dN/dE_p$ are the number distributions of neutrinos and relativistic protons, respectively.   This expression is obtained by integrating equation~\ref{eqn:L_nu} over the lifetime of a transient. As equation~\ref{eq:tpp} suggests that pp interactions are generally efficient at the peak time, below we take the pion production efficiency $f_{\rm pp} = 1$. Assuming that accelerated protons follow a power law spectrum, $dN/dE_p\propto E_p^{-q}$, and that $\epsilon_{\rm rel}$ fraction of the shock power is deposited into relativistic particles as described in equation~(\ref{eq:Erel}),  equation~(\ref{eqn:Phi_nu1}) can be rewritten as
\begin{equation}
    \Phi(E_\nu) = \frac{{\cal R}_0}{8\,\pi}\,\epsilon_{\rm rel}\,{\cal E}_{\rm opt}\,F_q\,\int dz\,\frac{c\,f(z)}{H(z)} \,(1+z)^{-q}
\end{equation}
with a prefactor 
\begin{equation}
\label{eqn:Phi_nu2}
F_q \equiv \begin{cases} (q-2)\,\left( {20\,E_\nu}/{E_{p,\rm min}}\right)^{2-q}\, &\mbox{if } q > 2 \\ 
\left(\log\,(E_{\rm max}/E_{\rm p, min})\right)^{-1} & \mbox{if } q = 2 \end{cases} 
\end{equation}
accounting for the integrated proton energy above  $E_{\rm p, min}$. 
For $E_{\rm p, min}\approx 1$~GeV, $E_{\rm max} \sim 10$~PeV, and $E_\nu \sim 100$~TeV, $F_q \approx 6.2\times 10^{-2}$, $1.1\times10^{-2}$, and $1.2\times 10^{-3}$ for $q = 2, \, 2.2,\, 2.4$, respectively.  The energy power-law index $q$ is equal to the momentum distribution index $q$ ($dN/dp \propto p^{-q}$) for relativistic particles ($E_p \simeq pc$), such that values $q \simeq 2-2.4$ are motivated both by the theory of diffusive particle acceleration and direct observations of novae (\S\ref{sec:novae}, Fig.~\ref{fig:Indrek}).   
 
The neutrino flux as measured by IceCube is $\Phi_{\nu_ + \bar{\nu}} \approx (4-6)\times10^{-8}\,\rm GeV\,cm^{-2}\,s^{-1}\,sr^{-1}$ at $E_\nu=100\,\rm TeV$ \citep{Stettner19, Schneider19}. To meet the observed diffuse neutrino flux, equation~\ref{eqn:Phi_nu2} poses a lower limit to ${\cal E}_{\rm opt}\,{\cal R}_0$ for given $\epsilon_{\rm rel}$ and $q$, following the argument connecting optical emission to non-thermal emission (eq.~\ref{eq:Erel} and surrounding discussion), 
\begin{eqnarray}
\label{eqn:limit}
\epsilon_{\rm rel} {\cal E}_{\rm opt} {\cal R}_0  &\approx& 5 \times 10^{44}\,\left(\frac{F_q}{F_q(q=2)}\right)^{-1}\,\left(\frac{\xi}{3}\right)^{-1}  \\ \nonumber
&\times&\left(\frac{\Phi_\nu^{\rm tot}}{4\times 10^{-8}\,\rm GeV\, cm^{-2}\, s^{-1}\, sr^{-1}}\right)\rm erg\,Mpc^{-3}\,yr^{-1}
\end{eqnarray}
where $\xi \equiv [\int dz f(z)H(z)^{-1} (1+z)^{-q}] / t_H$ (as first defined in \citet{1999PhRvD..59b3002W} with $q = 2$), $t_H = \int dz H(z)^{-1}(1+z)^{-1}$ is the age of the universe, and $\xi \approx 3$ for a star-forming history-like $f(z)$.

Figure~\ref{fig:energyrate_Emax} compares the maximum proton energy energy injection rate ${\cal E}_{\rm opt}\,{\cal R}_0$ of various transients derived in $\S\ref{subsec:derivedTransientQuantities}$ and the lower limit assuming $q = 2$, $\epsilon_{\rm rel} = 1$ and $\epsilon_{\rm rel} = 0.01$ (the latter as inferred from applying the calorimetric technique to novae; $\S\ref{sec:novae}$; Fig.~\ref{fig:Indrek}).  Fig.~\ref{fig:energyrate_Emax} shows that although a wide range of hypothesized shock-powered transients can accelerate ions to sufficient energies to explain the IceCube background, their neutrino production rates typically fall-short by $\gtrsim 2-4$ orders of magnitude in the favored case $\epsilon_{\rm rel} = 0.01$.

Finally, note that we have estimated neutrino production from proton-proton interaction. Nuclei with mass number $A>1$ lose energy both by fragmentation and pion production \citep{1994A&A...286..983M}, with the latter dominating above $\sim 1\,\rm TeV/A$ \citep{Krakau_2015}. 
Comparing to a proton, a nucleus with charge number $Z$ may gain $Z$ times more energy from the same acceleration zone (eqn.~\ref{eq:Emax}), though the energy per nucleon and hence the energy of their neutrino products is lower by a factor of $\sim Z/A\sim 1/2$ (see \citealt{2015JCAP...06..004F} for a comparison of neutrino production from Ap and pp interaction). The inelastic cross section of nuclei-proton interaction scales roughly by $A^{-1/3}$ \citep{2002cra..book.....S}, which allows efficient pion production at the peak epoch for most nuclei (eqn.~\ref{eq:tpp}). Nuclei-nuclei interaction (AA) would further complicate the secondary spectra comparing to Ap or pp interaction \citep{2012ApJ...750..118F}.  On the other hand, as the giant dipole resonance occurs at a lower energy with a larger cross section comparing to the photopion interaction (eqn.~\ref{eqn:t_pgamma}), photodisintegration may dominate over hadronuclear interaction and affect   neutrino production. A detailed computation of the competing processes is however beyond the scope of this work.   

\subsection{Propagation to Earth: Satisfying the Gamma-Ray Background Constraints}
\begin{figure}[htb!]
\includegraphics[width= \linewidth] {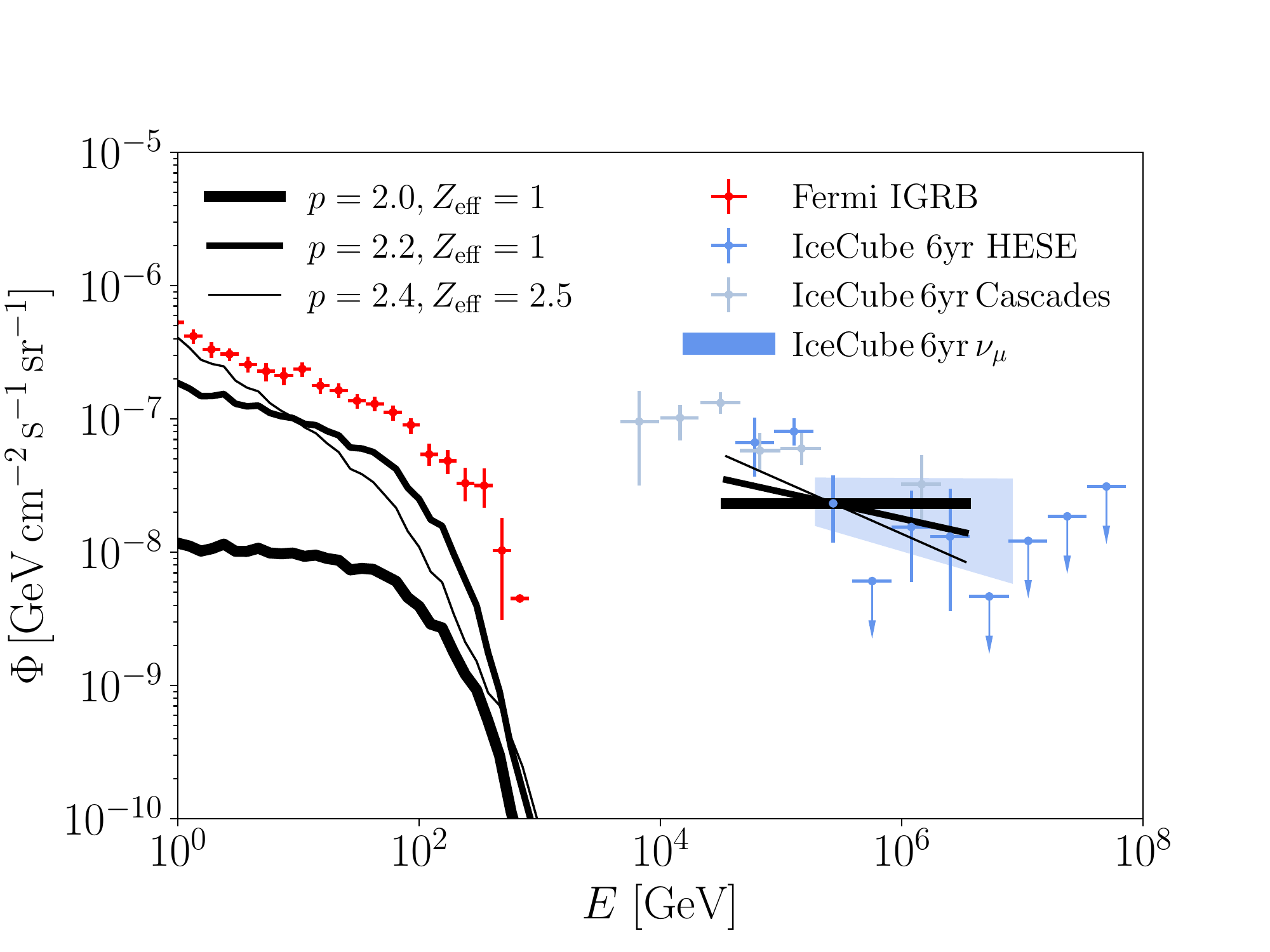} 
\caption{\label{fig:sed}
Diffuse neutrino and $\gamma$-ray fluxes from shock-powered transients comparing to the {\it Fermi}-LAT isotropic $\gamma$-ray background (IGRB)  \citep{2015ApJ...799...86A} and the diffuse neutrino flux of the high-energy starting events \citep{Schneider19}, $\nu_\mu$ events \citep{2016ApJ...833....3A}, and cascade events \citep{2020arXiv200109520I} measured by the IceCube Observatory.    The fluxes are computed by integrating the neutrino and $\gamma$-ray emission over a Type II SLSNe-like light curve (as in Figure~\ref{fig:L_t}), and summing over a source population that follows the star formation history. The grey solid curves assume injected spectral index $q = 2.0,\,2.2,\,2.4$ (from thick to thin) and an effective CSM charge number $Z_{\rm eff} = 1,\,1,\,2.5$, respectively. The fluxes are normalized by the IceCube HESE observation at 270~TeV. When taking a peak luminosity $L_{\rm opt}=10^{44}\,\rm erg\,s^{-1}$ and  $\epsilon_{\rm rel}=1\%$, the normalization corresponds to a local source rate of ${\cal R} =1,\, 17,\, 220\times 10^5\,\rm Gpc^{-3}\,yr^{-1}$  in each  scenario.       
}
\end{figure}

The flux of the diffuse neutrino background observed by the IceCube Observatory \citep{2016ApJ...833....3A, 2020arXiv200109520I} is comparable to that of the {\it Fermi}-LAT IGRB around $\sim 100$~GeV \citep{2015ApJ...799...86A}.  To avoid over-producing the IGRB, neutrino sources are suggested to be ``hidden" \citep{2001APh....15...87B}, being opaque to  1-100~GeV $\gamma$-rays or with hard $\gamma$-ray spectral index $\lesssim 2.1-2.2$ \citep{2016PhRvL.116g1101M}.

Around the peak time of a shock-powered transient when most of the high-energy neutrinos are produced, a significant fraction of GeV $\gamma$-rays may be attenuated due to the Bethe-Heitler process, depending on the charge number of the CSM (see \S\ref{subsec:gammaEscape}; also mentioned by \citealt{2017MNRAS.470.1881P, Murase+19}). Later the CSM becomes optically thin to GeV $\gamma$-rays, but proton-proton interaction is weaker and the shock power is lower. Shock-powered transients therefore emit much less energy in high-energy $\gamma$-rays than in high-energy neutrinos (see bottom panel of Fig.~\ref{fig:L_t}). 

To investigate whether these partially $\gamma$-ray dark sources satisfy the IGRB constraints, we evaluate the diffuse $\gamma$-ray and neutrino fluxes from shock-powered transients. The emission from an individual source is calculated as in Section~\ref{subsec:gammaEscape}, assuming an effective CSM charge number $Z_{\rm eff}$ and a proton spectrum with index $q$ and power $L_{\rm rel}\approx L_{\rm opt}\,\epsilon_{\rm rel}$.  The diffuse neutrino flux is then obtained by integrating the emission over a source life time and the evolution history of the source population following equation~\ref{eqn:Phi_nu1}. The diffuse $\gamma$-ray flux is computed by numerically propagating $\gamma$-rays from sources to the earth with Monte Carlo simulations. For the computation we adopt the extragalactic background light (EBL) model from \citealt{2011MNRAS.410.2556D} and an extragalactic magnetic field of $10^{-15}$~G on Mpc scales \citep{10.1093/mnrasl/sls026}.

Figure~\ref{fig:sed} present two benchmark scenarios, with $q=2,0,\,2.2,\,2.4$ and $Z_{\rm eff}=1,\,1,\,2.5$, respectively.  The fluxes are normalized to the IceCube high-energy starting event (HESE) data point at $\sim 270$~TeV. Both scenarios would over-produce the IGRB had the  source been transparent to $\gamma$-rays, but are safely below the IGRB due to the attenuation by the ejecta. 

\subsection{Requirements to Match the Neutrino Background}

Although shock-powered transients are promising as gamma-ray-dark sources, the known classes of transients we have considered come up several orders of magnitude short in terms of their energetic production (Fig.~\ref{fig:energyrate_Emax}).  To reproduce the overall normalization of the neutrino background, the scenarios shown in Fig.~\ref{fig:sed} require a hypothesized transient with $E_{\rm opt} = 5\times 10^{50}$~erg   and a particle acceleration efficiency $\epsilon_{\rm rel}=1\%$ with a (optimistic) local rate of ${\cal R}_0=1,\,17,\,220\times10^5\,\rm Gpc^{-3}\,yr^{-1}$ for $q=2, \,2.2,\,2.4$   respectively.  

In other words, we require some transient which is as frequent as core collapse supernovae, but emits $\gtrsim 50$ times the optical fluence.  Stated more precisely, we require a transient (or sum of transients) which obey  
\begin{eqnarray}
\left(\frac{\mathcal{R}_0}{10^{5}\,\rm Gpc^{-3}yr^{-1}}\right)\left(\frac{E_{\rm opt}}{5\times10^{50}\,\rm erg}\right)\left(\frac{\epsilon_{\rm rel}}{0.01}\right) \sim 1 
\label{eq:constraint}
\end{eqnarray}
for $q\approx 2$. Larger values of q would require even larger values of $\mathcal{R}_0$ and/or $E_{\rm opt}$ as described by equation~\ref{eqn:limit}.

\section{Summary and Conclusions}
\label{sec:conclusions}

We have introduced a simple technique for combining the observed properties of non-relativistic optical transients to their maximal high-energy neutrino and gamma-ray outputs in order to constrain their contributions to the IceCube and {\it Fermi} backgrounds.  Our conclusions may be summarized as follows: 
\begin{itemize}

\item A large number of optical transients could in principle be shock-powered (Table~\ref{tab:classes}), even if the direct signatures of shock interaction (e.g.~emission lines) are hidden at early times.  Despite a diversity of dynamics and geometry, a generic feature of their behavior is a shock which propagates outwards in time from high to low optical depths through some  medium which covers a fraction of the total solid angle (Fig.~\ref{fig:cartoon}).  

\item The condition for the creation of a collisionless shock capable of accelerating relativistic ions is similar to that for the escape of optical radiation.  Thus, relativistic particle acceleration commences around the time of optical maximum, $t_{\rm pk}$, which for most transients is also the epoch at which the majority of the optical radiation energy is released.

\item The calorimetric technique makes use of the fact that at the epoch $\sim t_{\rm pk}$ the cooling time of both thermal and non-thermal particles (via free-free emission and p-p interactions, respectively) is generically short compared to the expansion time (eqs.~\ref{eq:tcool},\ref{eq:tpp}).  As a result, the energy radiated by non-thermal ions in high-energy neutrinos and gamma-rays is directly proportional to the transient's shock-powered optically energy.  The proportionality constant is the ion acceleration efficiency, $\epsilon_{\rm rel}$ (eq.~\ref{eq:Erel}).  

\item Observations of correlated optical and gamma-ray emission in classical novae (e.g.~\citealt{Li+17,Aydi+20}) enable a proof-of-principle application of the calorimetric technique which probes the properties of ion acceleration at radiative internal shocks under physical conditions similar to those which characterize more luminous extragalactic transients.  The ratio of optical to gamma-ray luminosities reveal ion acceleration efficiencies $\epsilon_{\rm rel} \sim 1\%$, while an analysis of the gamma-ray spectra are consistent with relatively flat injected ion spectra ($q \lesssim 2.4$) and energy cut-off $E_{\rm max} \sim 30$ GeV (Fig.~\ref{fig:Indrek}).  

\item We make a simple estimate for the maximum particle energy accelerated at radiative shocks (eqs.~\ref{eq:Emax1},\ref{eq:Emax2}), which unlike most previous studies accounts for the thin radial extent of the downstream region due to radiative compression.  Applying this formalism to gamma-ray data from classical novae ($E_{\rm max} \gtrsim 30$ GeV) require magnetic amplification at the shocks, $\epsilon_{\rm B} \sim 10^{-2}$.  Assuming a similar magnetic field amplification factor in the shocks of extragalactic transients, we find that many exceed the threshold $E_{\rm max} \gtrsim 10^{15}$ eV needed to generate neutrinos above 50~TeV (Fig.~\ref{fig:energyrate_Emax}) and hence contribute to the IceCube diffuse neutrino flux. 

\item Due to the high Bethe-Heitler optical depth of the ejecta at the epoch of peak neutrino fluence $t_{\rm pk}$ (Fig.~\ref{fig:tau_gamma}), we confirm previous suggestions (e.g.~\citealt{2017MNRAS.470.1881P, Murase+19}) that shock-powered transients can in principle serve as gamma-ray-hidden  neutrino sources (Fig.~\ref{fig:sed}) consistent with the non-blazar {\it Fermi}-LAT background.

\item Using the inferred energetics and volumetric rate of each class of transient we calculate its maximal neutrino output, derived under the assumption that 100\% of its optical radiation is powered by shocks.  Even in this most-optimistic case, we find that the classes of known optical transients we have considered are insufficient to explain the IceCube background (Fig.~\ref{fig:energyrate_Emax}) unless they produce a hard proton spectrum with index $q\sim2$ or lower. With $q>2.2$ they individually fall short by $\gtrsim2-3$ magnitudes if we adopt a value $\epsilon_{\rm rel} = 1\%$ calibrated to classical novae.  Even making the optimistic assumption that all core collapse supernovae in the universe are 100\% shock-powered, the normalization of the background is achieved only in the unphysical case $\epsilon_{\rm rel} \sim 1$.

\item The most promising individual sources are TDEs, but whether the light curves of these sources is powered by shocks (e.g.~\citealt{Piran+15}) or reprocessed X-rays from the inner accretion flow (e.g.~\citealt{Metzger&Stone16}) is hotly-debated. It has been suggested that the TDE rate decreases with redshift \citep{2016MNRAS.461..371K}, in which case the neutrino flux would be lower than our estimation based on the evolution of the cosmic star formation rate. For reference, $\xi$ (eq.~\ref{eqn:limit}) decreases from 2.8 for a star-formation evolution to 0.6 for a uniform source evolution.  

Interestingly, \citet{Stein+20} recently reported that an IceCube neutrino alert event arrived in the direction of a radio-emitting TDE around $\sim 180$~days after discovery (see also \citealt{Murase+20,Winter&Lunardini20}). The probability of a coincidence by chance is $0.2-0.5\%$. No $\gamma$-ray signal was detected by the {\it Fermi}-LAT, implying that $\gamma$-rays may have been attenuated by the UV photosphere  \citep{Stein+20}, similar to what we suggest in this work.  However, our model would predict that neutrinos arrive around the peak time of the optical/UV emission of a TDE, which was around a month after discovery for this event.

\item Although we have focused on radiative shocks, which we have shown to characterize shock-powered optical transients near peak light, for a
lower CSM density$-$such as encountered at later times and larger radii$-$the shocks will instead be adiabatic and our calorimetric argument will break down.  However, this is unlikely to significantly
change our conclusions because, for most CSM density profiles, the total shock-dissipated energy is still dominated by early times, when the shocks are radiative.  Furthermore, the efficiency of relativistic particle acceleration at non-relativistic, quasi-perpendicular adiabatic shocks may be even lower than in radiative shocks with the same upstream magnetic field geometry due to the effects of thin-shell instabilities on the shape of the shock front (\citealt{Steinberg&Metzger18}).

\item Several of the transient classes considered in our analysis (e.g., FBOTs) have only been discovered and characterized in the past few years.  We therefore cannot exclude that another class of optical transients will be discovered in the future which is more promising as a background neutrino source.  However, given the stringent requirement on the product of volumetric rate and optical energy fluence placed by equation (\ref{eq:constraint}) to match the IceCube flux, it is hard to imagine that recent or existing synoptic surveys (e.g.~ZTF, PanSTARRs) have missed such events completely.  One speculative exception would be a source class restricted to the high redshift universe, in which case the greater sensitivity and survey speed of the Vera C.~Rubin Observatory would be required for its discovery.  One may also speculate about the existence of a class of optically-dark but infrared bright transients missed by previous surveys (e.g.~\citealt{Kasliwal+17}).       
\end{itemize}

\begin{deluxetable*}{cccccccc}
\tablecaption{Observed Properties of Extragalactic\tablenotemark{$\dagger$} Transients\label{tab:classes}}
\tablewidth{700pt}
\tabletypesize{\scriptsize} 
\tablehead{
\colhead{Source} & \colhead{$\mathcal{R}_0$\tablenotemark{\scriptsize s}} & 
\colhead{log$_{10}L_{\rm pk}$} & \colhead{$t_{\rm pk}$} & 
\colhead{$\bar{v}_{\rm ej}$} & \colhead{log$_{10}E_{\rm opt}$\tablenotemark{\scriptsize t}} & 
\colhead{$Z$\tablenotemark{\scriptsize u}} & \colhead{Shock} \\ 
\colhead{} & \colhead{($\rm Gpc^{-3}\,yr^{-1}$)} & \colhead{($\rm erg\,s^{-1}$)} & \colhead{(d)} & 
\colhead{(10$^{3}$ km s$^{-1}$)} & \colhead{(erg)} & \colhead{}  &  \colhead{Powered?} 
} 
\startdata
Novae & $(1-5)\times 10^{8}$ & 37--39 & 3 & $0.5-3$ & $43.5-44.5$  & 1 & Y\tablenotemark{\scriptsize a} \\
LRN & $10^{5.5}-10^{6.4}$ \tablenotemark{\scriptsize b} & 39-41  & 40--160 & $0.2-0.5$   & $45-46$  & 1  & ?\tablenotemark{\scriptsize c} \\
SLSNe-I & 10--100 \tablenotemark{\scriptsize d} & 43.3--44.5 \tablenotemark{\scriptsize e} & 30--50  & $5-10$ & $50-51$ & 8 & ? \\
SLSNe-II & 70--300 \tablenotemark{\scriptsize f} &  43.6--44.5 & 31--36 & $5-10$   & $50-51$  & $1$ &Y \\
SNe-IIn \tablenotemark{\scriptsize g} & $3000$ \tablenotemark{\scriptsize h} &  42--43.7 & 20--50 & $5$   &  $49-50$  & 1 &Y \\
CCSNe &  $7\times10^4$ \tablenotemark{\scriptsize i} & 41.9--42.9 & 7--20 \tablenotemark{\scriptsize j} & 3  & $48-49$  & 1,8 & ?? \\
TDE & 100--1000 \tablenotemark{\scriptsize k} & 44--45 \tablenotemark{\scriptsize l}  & 40--200 \tablenotemark{\scriptsize m} & $5-15$  & $51-52$ & 1  &? \\
FBOT & $\sim 4800-8000$ \tablenotemark{\scriptsize n}& $\sim 43$ &  $4-12$\tablenotemark{\scriptsize n} &  $6-30$   & $48.5-49.5$  & ? & ? \\
Lum.~FBOT & $\sim 700-1400$\tablenotemark{\scriptsize o} & $\sim 44$ & 1-5\tablenotemark{\scriptsize p} & $6-30$ \tablenotemark{\scriptsize q}  & $49.5-50.5$  & 1 & ? \\
Type Ia-CSM & 300-3000\tablenotemark{\scriptsize r} & $\sim$ 43 & 20 & 10 & 49 & 6-8 & Y
\enddata
\tablenotetext{\dagger}{LRN and Novae are also frequent Galactic transients.  }
\tablenotetext{a}{\citealt{Li+17,Aydi+20}}
\tablenotetext{b}{\citealt{Kochanek14}}
\tablenotetext{c}{\citealt{MetzgerPejcha17}}
\tablenotetext{d}{\citealt{Quimby+13} at $z=0.17$ with $h=0.71$}
\tablenotetext{e}{\citealt{Inserra19} }
\tablenotetext{f}{ \citealt{Quimby+13} at $z=0.15$ with $h=0.71$}
\tablenotetext{g}{ \citealt{2011MNRAS.412.1522S, 2014ApJ...789..104O,   2019arXiv190605812N}}
\tablenotetext{h}{ Taken to be 8.8\% of the total CCSNe rate; \citealt{Li+11}}
\tablenotetext{i}{ \citealt{Li+11, 2014ApJ...792..135T}}
\tablenotetext{j}{ \citealt{2015MNRAS.451.2212G}} 
\tablenotetext{k}{ \citealt{2014ApJ...792...53V, 2014MNRAS.444.1041K, StoneMetzger16}.}
\tablenotetext{l}{ Blackbody fits to optical data, \citealt{vanVelzen18}}
\tablenotetext{m}{\citealt{Mockler+19}}
\tablenotetext{n}{\citealt{Drout+14} at $z \lesssim 0.65$ }
\tablenotetext{o}{ Taken to be 0.25\% of the CCSN rate at $z = 0.2$; \citealt{Coppejans+20}}
\tablenotetext{p}{ \citealt{Prentice+18},\citealt{Ho+20}}
\tablenotetext{q}{Width of the late-time emission features in AT2018cow; \citealt{Perley+19}. }
\tablenotetext{r}{Taken to be 0.1-1\% of the Type Ia rate; \citealt{Dilday+12}.}
\tablenotetext{s}{Local $z = 0$ volumetric rate of transient class.}
\tablenotetext{t}{Total radiated optical energy per event.}
\tablenotetext{u}{Average nuclear charge in ejecta.}

\end{deluxetable*}

\begin{deluxetable*}{ccccc}
\tablecaption{Derived Properties of Extragalactic\tablenotemark{$\dagger$} Transients\label{tab:classes_derived}}
\tablewidth{700pt}
\tabletypesize{\scriptsize}
\tablehead{
\colhead{Source} & \colhead{log$_{10}\,\dot {\cal E}_{\rm opt}$} & 
\colhead{$f_{\Omega,\rm min}$} &  
\colhead{$v_{\rm sh}$} & \colhead{$\log\,(E_{\rm max}/{\rm eV})$} \\ 
\colhead{} & \colhead{($\rm erg\,Mpc^{-3}\,yr^{-1}$)} & \colhead{} &  
\colhead{($10^3$~km s$^{-1}$)} &   \colhead{} 
} 
\startdata
Novae & 43.2 &   $4.0\times10^{-6}$ & 0.1 -- 3.0 & $<9-13.8$   \\
LRN   & 42.4 & $1.6\times10^{-3}$ & 0.1 -- 0.5 & $<9-8.1$  \\
SLSNe-I & 43.0 &  $1.3\times10^{-2}$   & 2.4 -- 10.0 &  14.0 -- 16.1  \\
SLSNe-II & 43.5 & $3.6\times10^{-2}$ & 3.3 -- 10.0 & 14.6 -- 16.1 \\
SN-IIn  & 43.5 & $5.2\times10^{-3}$ & 0.9 -- 5.0 & 11.1 -- 14.9  \\
CCSNe  & 43.8 &  $4.8\times10^{-2}$   & 1.1 -- 3.3 & 11.2 -- 14.0  \\
TDE   & 45.0 & $4.9\times10^{-3}$ & 2.5 -- 15.0 & 14.7 -- 17.4 \\
FBOT & 43.4 & $1.0\times10^{-3}$ & 3.0 -- 30.0 & 14.9 -- 18.3 \\
Lum.~FBOT & 43.6 & $2.4\times10^{-2}$  & 8.7 -- 30.0 & 16.5 -- 17.4 \\
Type Ia-CSM & 43.5 & $1.6\times10^{-2}$ & 2.5 -- 10.0 & 13.9 -- 15.7 \\ 
\enddata

\tablenotetext{\dagger}{LRN and Novae are also frequent Galactic transients.  }
\end{deluxetable*}

\acknowledgments
Support for KF was provided by NASA through the NASA Hubble Fellowship grant \#HST-HF2-51407 awarded by the Space Telescope Science Institute, which is operated by the Association of Universities for Research in Astronomy, Inc., for NASA, under contract NAS5-26555.  BDM was supported in part by the Simons Foundation (grant \# 606260); by the National Science Foundation (grant \#AST-1615084); and by NASA Hubble Space Telescope Guest Investigator Program (grant \#HST-AR-15041.001-A).  EA and LC acknowledge NSF award AST-1751874, NASA award 11-Fermi 80NSSC18K1746, and a Cottrell fellowship of the Research Corporation. IV acknowledges support by the Estonian Research Council grants IUT26-2 and IUT40-2, and by the European Regional Development Fund (TK133).

\bibliographystyle{aasjournal}
\bibliography{ref}

\end{document}